\documentclass[a4paper,11pt]{article}
\usepackage{jcappub} 
\usepackage{xspace}
\usepackage[utf8]{inputenc}
\usepackage{textgreek}
\usepackage{graphicx}
\usepackage{verbatim}
\usepackage{rotating} 
\usepackage{xspace}
\usepackage{mathtools}
\usepackage{amsmath}
\usepackage{longtable}
\usepackage{comment}
\usepackage{multirow}
\usepackage{caption}
\usepackage{subcaption}
\usepackage{appendix}
\usepackage{afterpage}
\usepackage{changepage}
\usepackage{graphicx} 
\usepackage{verbatim}
\usepackage[utf8]{inputenc}
\usepackage{natbib}
\usepackage{amsmath}
\usepackage{amsfonts}
\usepackage{mathtools}
\usepackage{multirow}
\usepackage{makecell}
\usepackage{changepage}
\usepackage{amssymb}
\usepackage{gensymb}
\usepackage{acronym}

\def\flat{\textit{Fermi}/LAT\xspace}

\def\hess{H.E.S.S.\xspace}
\graphicspath{ {./images/} }

\title{Prospects for annihilating dark matter from M31 and M33 observations with the Cherenkov Telescope Array}

\author{Miltiadis Michailidis$^1$} 
\author{Lorenzo Marafatto$^2$}
\author{Denys Malyshev$^1$}
\author{Fabio Iocco$^{3,4}$}
\author{Gabrijela Zaharijas$^2$}
\author{Olga Sergijenko$^{5,12,13}$}
\author{Maria Isabel Bernardos$^6$}
\author{Christopher Eckner$^{2,7}$}
\author{Alexey Boyarsky$^8$}
\author{Anastasia Sokolenko$^{9,10,11}$}
\author{Andrea Santangelo$^1$}

\affiliation{$^{1}$ Institut f{\"u}r Astronomie und Astrophysik T{\"u}bingen, Universit{\"a}t T{\"u}bingen, Sand 1, D-72076 T{\"u}bingen, Germany}

\affiliation{$^{2}$ Center for Astrophysics and Cosmology, University of Nova Gorica, Vipavska 13, SI-5000
Nova Gorica, Slovenia}
\affiliation{$^{3}$ Dipartimento di Fisica ”Ettore Pancini”, Università degli studi di Napoli
“Federico II”, Complesso Univ. Monte S. Angelo, I-80126 Napoli, Italy}

\affiliation{$^{4}$ INFN - Sezione di Napoli, Complesso Univ. Monte S. Angelo, I-80126 Napoli, Italy}
\affiliation{$^{5}$ Astronomical Observatory, Taras Shevchenko National University of Kyiv, Kyiv, Ukraine}
\affiliation{$^{6}$ Dipartimento di Fisica e Astronomia dell'Università and Sezione INFN, Padova,
Italy, Padova, Italy}
\affiliation{$^{7}$ LAPTh, CNRS,  USMB, F-74940 Annecy, France}
\affiliation{$^{8}$ Lorentz Institute, Leiden University, Niels Bohrweg 2, Leiden, NL-2333 CA, The Netherlands}
\affiliation{$^{9}$ University of Chicago, Kavli Institute for Cosmological Physics, Chicago, IL 60637, USA}
\affiliation{$^{10}$ Fermi National Accelerator Laboratory, Theoretical Astrophysics Group, Batavia, IL 60510, USA}
\affiliation{$^{11}$ Institute of High Energy Physics, Austrian Academy of Sciences, Nikolsdorfergasse 18, 1050 Vienna, Austria}
\affiliation{$^{12}$ Main Astronomical Observatory of the National Academy of Sciences of Ukraine, Zabolotnoho str., 27, 03143, Kyiv, Ukraine}
\affiliation{$^{13}$ Space Technology Centre, AGH University of Science and Technology, Aleja Mickiewicza, 30, 30-059, Kraków, Poland}

\abstract{
M31 and M33 are the closest spiral galaxies and the largest members (together with the Milky Way) of the Local group, which makes them interesting targets for indirect dark matter searches. In this paper we present studies of the expected sensitivity of the Cherenkov Telescope Array (CTA) to an annihilation signal from weakly interacting massive particles from M31 and M33. We show that a 100~h long observation campaign will allow CTA to probe annihilation cross-sections up to $\langle\sigma\upsilon\rangle\approx5\cdot10^{-25}~\mathrm{cm^{3}s^{-1}}$ for the $\tau^{+}\tau^{-}$ annihilation channel (for M31, at a DM mass of 0.3~TeV), improving the current limits derived by HAWC by up to an order of magnitude. 
We present an estimate of the expected CTA sensitivity, by also taking into account the contributions of the astrophysical background and other possible sources of systematic uncertainty.
We also show that CTA might be able to detect the extended emission from the bulge of M31, detected at lower energies by the \flat. 
}

\begin{document}
\maketitle
\section{Introduction}\label{zero}

Cosmological and astrophysical observations of diverse nature suggest that the majority of the matter in the Universe consists of a non-electromagnetically interacting component, often referred to as Dark Matter (DM), see e.g. \citet{2005PhR...405..279B, pdg20}. Despite the DM density having been measured with a great accuracy to be $\Omega_{DM} h^2 = 0.11933 \pm 0.00091$~\citep{2020}, little is known about its very nature. 

Whereas different scenarios with regards to the nature and origin of DM that have been proposed by physicists throughout the years --such as for instance Primordial Black Holes-- have not been entirely ruled out at the moment (~\citep[see e.g.][]{pbh_review}), yet ample data evidence keep holding around the fact that it is more likely that the DM nature is non-baryonic, thus requiring physics beyond the Standard Model (SM)~\citet{2005PhR...405..279B, pdg20}.
Indeed, many SM extensions proposed to date naturally include a DM candidate, namely a particle complying with all astrophysical and cosmological requirements, and produced in the right abundance in the early Universe, see e.g.~\citet{pdg20} for a recent review of such candidates.

Within the broadly considered SM extensions providing DM candidates, the Weakly Interacting Massive Particles (WIMPs) are one of the most widely explored in particle and astroparticle physics. MeV -- TeV mass scale self-annihilating WIMPs with a weak-scale cross-section (DM-particles velocities averaged annihilation cross-section $\langle\sigma \upsilon\rangle_{th}=3\cdot 10^{-26}$cm$^3$s$^{-1}$) can naturally produce the observed abundance of the DM as a result of thermal freeze-out in the early Universe, see~\citep{1977PhRvL..39..165L,2008PhRvL.101w1301F}, and \citet{2013arXiv1301.0952P,BAER20151}.

If WIMPs constitute the entirety of the DM, their annihilation into the SM particles with the consequent production of photons~\citep[see e.g.][for a review]{Cirelli_2011} makes WIMPs good candidates for indirect searches of the annihilation signal from certain DM-dominated objects. The produced photons are expected to have a hard spectrum which continues up to WIMP's mass. While the exact shape of the spectrum depends on the type of SM particles into which WIMPs primarily annihilate (``annihilation channel''), the maximum of the spectral energy density is located in the TeV band for a TeV-scale WIMP. This makes the very high energy (VHE) band an important window for indirect WIMP-DM searches.

The TeV band is currently being explored by several Imaging Atmospheric Cherenkov Telescopes (IACTs). These facilities utilise Cherenkov radiation from the secondary particles produced in interactions of primary cosmic rays with the atmosphere to detect and characterise the properties of the incident primary particle. Currently, major operational IACTs are \hess (located in Southern Hemisphere), MAGIC, and VERITAS (both -- Northern Hemisphere).

During the last decade these telescopes performed a number of dedicated WIMP DM search campaigns in the TeV band. These include a dedicated multi-year campaign for the search of the annihilation of WIMPs close to the Galactic Center (GC) region with \hess~\citep{hess_gc_linelike,hess_gc,hess1_gc,hess_dm_limits_review};
individual and joint multi-facility campaigns on nearby dwarf spheroidal galaxies (dSphs)~\citep{Aliu_2009,Acciari_2010,2016JCAP...02..039M,zitzer2017veritas,yapici2017dark,hess_wlm,oakes2019combined},
DM annihilation searches in nearby galaxy clusters~\citep{clusters_hess} and searches for clumps of DM in our galaxy~\citep{hess_ufos_icrc, hess_ufos}. For a complete report of all observations performed by current IACTs see \citep{2021arXiv211101198D}.

At somewhat higher energies ($\gtrsim 10$~TeV) DM searches are extensively performed by high-altitude broad field of view instruments such as e.g. ARGO-YBJ~\citep{argo_def} (currently decommissioned), HAWC~\citep{hawc_def} and most recently the LHAASO~\citep{lhaaso_def} observatory. The tightest constraints on the parameters of annihilating DM provided by these facilities arise from the non-detection of a DM annihilation signal in the MW halo~\citep{hawc_halo}, dSphs~\citep{hawc_dsphs}, DM sub-halos~\citep{hawc_clumps} and nearby galaxies~\citep{hawc_m31}.

In the GeV-TeV band, the WIMPs' properties are constrained dominantly by the space-based missions, e.g., \flat~\citep{lat_def}. The primary targets for the searches in this band were dwarf spheroidals~\citep{fermi_dsphs,fermi_dsphs1,fermi_dsphs2}, galaxy clusters~\citep{Aleksi__2010,Arlen_2012,fermi_clusters1,fermi_clusters2,fermi_clusters}, Galactic Center observations~\citep[see e.g.][and references therein]{fermi_gc_dm1,fermi_gc_dm2}, nearby galaxies~\citep{fermi_m31_dm,2019PhRvD..99l3027D} and DM sub-halos~\citep{https://doi.org/10.48550/arxiv.1109.5935,https://doi.org/10.48550/arxiv.1509.00085,Coronado_Bl_zquez_2019_1,Coronado_Bl_zquez_2019_2}.

Despite enormous dedicated efforts, the state-of-the-art WIMP DM searches only marginally approach the thermal annihilation cross-section scale. The best limits are obtained for WIMP masses $\lesssim 0.1$~TeV, which are based on the joint-analysis of the observational data from 27 dSphs by \flat~\citep{fermi_dsphs}. For $b\bar{b}$ and $\tau^+\tau^-$ annihilation channels in this mass range, the derived limits are by an order of magnitude better than the thermal cross-section~\citep[see however][]{fermi_dsphs_uncert}. For higher DM masses, the tightest constraints resulted from a dedicated multi-year 254~h long \hess observational campaign on the GC. For a preferable DM profile, \citet{Abdallah_2016} have shown that the obtained \hess limits can reach the thermal cross-section for the $\tau^+\tau^-$ annihilation channel and WIMP masses of the order of $\sim 1$~TeV, while at higher masses the derived limits are quickly degrading.

The gap between the sensitivity of current-generation instruments and the required sensitivity to probe the thermal annihilation cross-section in a broad portion of the WIMP parameter space offers ample opportunities for next-generation facilities to push forward the frontiers in indirect DM searches. Some of these facilities (e.g., LHAASO) already produced first results and are performing DM-dedicated campaigns~\citep[see e.g.][]{lhaaso_dsphs,lhaaso_halo}, while others (e.g., Cherenkov Telescope Array (CTA)) are still in the construction phase.

The CTA will be composed of two sites, one in the Northern (La Palma, Canary Islands, Spain) and one in the Southern Hemisphere (Paranal Observatory, Chile), which will enable observations to cover the entire Galactic plane and a large fraction of the extra-galactic sky~\citep[see e.g.][]{cta_def}. The arrays will include three different telescope sizes to maximize the energy range of the instrument (from 20 GeV to more than 300 TeV). With more than 100 telescopes in the Northern and Southern Hemispheres combined, in the next decade, the CTA will be the largest ground-based IACT $\gamma$-ray observatory in the world. The CTA will have an order of magnitude higher effective area and broader field of view than the current generation of IACTs~\citep{cta_book}. This makes CTA one of the best instruments for indirect DM searches at TeV energies. 

Present-day indirect DM searches are focused on several classes of objects, which include such DM-dominated objects as dwarf spheroidal galaxies; clusters of galaxies, or the MW's Galactic Center. As a viable alternative to these commonly considered objects, we consider studies of the annihilation DM signal from nearby spiral galaxies (i.e., M31 and M33). The DM search in such galaxies (M33) had been previously performed in 2008 with the Whipple 10~m $\gamma$-ray telescope~\citep{2008ApJ...678..594W} and recently by HAWC~\citep{hawc_m31}, towards M31 resulting in competitive to other targets constraints. In what follows, we perform detailed studies to address the CTA potential to constrain the parameters of annihilating WIMP DM using observations of M31 and M33. We note also that M31 is the subject of a $\sim 150$~h long key science program. The proposed indirect DM search can additionally strengthen the scientific goals of that program.

Τhe paper is organized as follows. In section~\ref{II}, we describe the motivation for selecting M31 and M33 from all nearby spiral galaxies for this study. In this section we also quantitatively describe the expected signal from annihilating WIMPs as well as summarize details of astrophysics back/fore-ground emission relevant to the analysis. CTA data simulation and analysis are described in~Section~\ref{III}. In section~\ref{IV}, we report on the CTA's sensitivity to an annihilating WIMP signal for several considered annihilation channels. Special attention is devoted to an accurate treatment of uncertainties related to the astrophysical background, the lack of knowledge of the actual DM density distribution in the considered objects as well as the impact of instrumental systematic uncertainties. Finally, in section~\ref{V}, we shortly summarize the derived conclusions.

\section{Expected signal and Target selection}\label{II}
\subsection{DM annihilation signal}\label{signal}

WIMP annihilation with its antiparticle (that in many scenarios is the WIMP itself, a Majorana particle) leads to the production of SM particles. Depending on the type of the produced SM particles several annihilation channels (e.g. quark $b\bar{b}$, $t\bar{t}$, leptonic $\tau^+\tau^-$ , $\mu^+\mu^-$ or bosonic $W^+W^-$, $ZZ$ annihilation channels) can be contemplated. Annihilation/decay of the produced SM particles results in the emission of secondary photons, which can be detected with ground or space-based observatories.

The same DM annihilation process taking place in the early Universe is to be expected in all environments, and it will depend on the local DM density. 
In astrophysical objects with a given DM density distribution $\rho(r)$, the observed signal is therefore characterised by a spatial and spectral components~\citep[see for more details a review by][]{1998APh.....9..137B}

\begin{equation}
   \frac{d\Phi}{dE_{\gamma}d\Omega}=\frac{1}{8\pi}\cdot\frac{\langle\sigma\upsilon\rangle}{m^{2}_{\chi}}\cdot\left.\frac{dN_{\gamma}}{dE_{\gamma}}\right|_{i}\cdot\ \int\limits_{l.o.s.} \rho^{2}_{DM}(r(\ell),\Omega) d\ell
   \label{eq:annihil_flux}
\end{equation}
where $m_{\chi}$ is WIMP's mass and $i$ presents WIMPs primary annihilation channel.

The differential term $\frac{d\Phi}{dE_{\gamma}d\Omega}$ on the left side of this equation corresponds to the observed photon flux. The right-hand side can be thought of as a product of two factors: \textit{(i)} astrophysical, determined by DM density content in the object ($J$-factor)
\begin{align}
& dJ/d\Omega = \int\limits_{l.o.s.} \rho^{2}_{DM}(r(\ell),\Omega) d\ell   
\label{eq:j_factor}
\end{align}
 and expressed as the line of sight (l.o.s.) integral of the DM density squared within the solid angle $d\Omega$ of the observation. Where $\ell$ is the variable that parametrizes the l.o.s., and $r$ is the radial distance from the center of the selected target. \textit{(ii)} particle physics term $\left.\frac{dN_{\gamma}}{dE_{\gamma}}\right|_{i}$ presenting the final-state photon spectrum of one annihilation of DM particles annihilating via annihilation channel $i$. The remaining coefficients serve to account for the frequency of annihilation events (DM-particles velocity-averaged annihilation cross-section $\langle\sigma\upsilon\rangle$) and relate to the number of annihilation events (term $m_x^{-2}$). 

Among all possible annihilation channels, we focus here only on $b\bar{b}$ (``benchmark channel'' in what below), $\tau^+\tau^-$ and $W^+W^-$ channels as widely discussed representatives of the annihilation channels.

\begin{figure}[!h]
    \centering
    
    \includegraphics[width=0.8\textwidth]{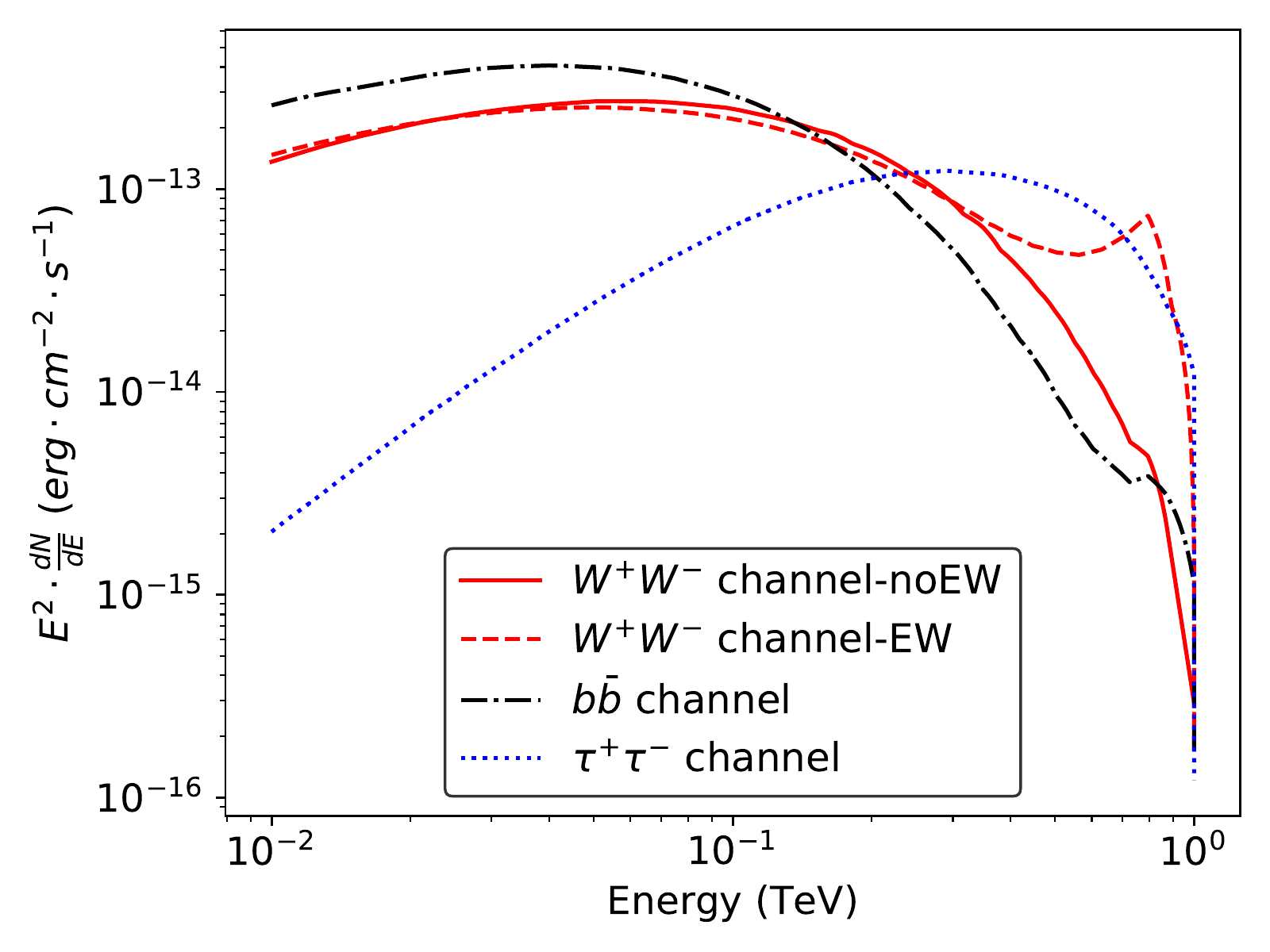}
    \caption{Expected 1~TeV mass WIMP DM annihilation spectrum in the M31 assuming the benchmark density model. The red, blue and black lines show the annihilation spectra in $W^+W^-$, $b\bar{b}$ and $\tau^+\tau^-$ channels.}
    \label{fig:spec_3cha}
\end{figure}

The typical spectral shape of the signal expected from the annihilation of DM with the mass $m_x=1$~TeV is shown in Fig.~\ref{fig:spec_3cha} which is based on~\citep{2011JCAP...03..051C} tables. We note that the spectrum of $W^+W^-$ annihilation channel can be substantially different at energies close to the $m_x$ with/without accounting for electroweak (EW) corrections~\citep{Ciafaloni_2011}.  For this channel, we explicitly present results corresponding to the spectrum obtained without EW corrections and to the spectrum which is based on the model-independent treatment of EW corrections. Similarly, one could employ the HDMSpectra\footnote{https://github.com/nickrodd/HDMSpectra} code introduced in \citep{2021JHEP...06..121B} which performs similar calculations to \citep{2011JCAP...03..051C} code aiming at constructing the obtained spectral shape from DM annihilation signal. 
This code computes DM annihilation spectra for DM masses above the EW symmetry breaking and all the way to the Planck scales, thus it is considered to considerably improve computed results for the $W^+W^-$ annihilation channel by accounting for all relevant EW corrections. However, to be conservative, we explicitly derive all the 
results for the $W^+W^-$ annihilation channel without accounting for EW corrections. Namely,  we did not include EW corrections for the $W^+W^-$ annihilation channel, since such corrections are model-dependent and account for a strong modification in the energy spectrum of DM particles with masses greater than the electroweak scale \citep{Ciafaloni_2011,2011JCAP...03..051C,2012JCAP...10E.001C}. In more detail, EW corrections are responsible for a slight enhancement of the lower energy regime of the spectrum due to the conversion of a small portion of high-energy particles to a substantial amount of lower-energy particles. In addition, forbidden final states are enabled, resulting in the presence of the whole population of stable particles in the final spectrum, regardless of the primary channel of annihilation. Finally, they are responsible for the perception of a model-dependent strong peak, with an energy value associated with the DM mass, in the DM annihilation spectrum through the $W^+W^-$ annihilation channel which determines all the constraints prevailing over the entire spectrum~\citep{2019JCAP...12..061V}. The impact of EW correction to the DM annihilation spectrum, through the $W^+W^-$ channel, is illustrated in Fig.~\ref{fig:spec_3cha} with the red-dashed line.

\subsection{Selected Targets}
\label{sec:target_selection}
M31 (Andromeda Galaxy) and M33 (Triangulum Galaxy) are the DM-dominated spiral galaxies closest to the Milky Way, which makes them potentially interesting targets for indirect searches of decaying or annihilating DM.  Located at distances of 
$778~\mathrm{kpc}$ \citep[M31, see e.g.] []{2004AJ....127.2031K} and $840~\mathrm{kpc}$ \citep[M33 ; ][]{1991ApJ...372..455F} these galaxies are among the best-studied objects in terms of DM density distribution.

The relative proximity of M31 and M33 galaxies allowed several dedicated studies of the DM profiles in these objects. The comprehensive list of DM profiles presented in the literature for these objects is given in Tab.~\ref{tab:profiles} and Tab.~\ref{tab:profiles2}. The Tables summarise the basic information on the galaxies (coordinates, distance, visibility from Southern or Northern CTA site) as well as parameters of DM density profiles. The last ones include density profile adopted in the corresponding study (isothermal~\citep[ISO;][]{1966AJ.....71...64K}, Navarro-Frank-White~\citep[NFW;][]{1997ApJ...490..493N}, Einasto \citep{1965TrAlm...5...87E}), Burkert~\citep{1995ApJ...447L..25B}, see also see Appendix~\ref{appendA}), parameters of the profiles (characteristic radius $r_s$ and density $\rho_s$) and the bibliographic reference for the work reporting the corresponding profile.

\subsection{Benchmark DM density profiles}
\label{sec:bench_models}
Demonstrating a good agreement at large distances from M31 and M33 centers, dark matter density profiles are still rather uncertain closer to the centers of these objects. In what follows, we select Einasto (for M31) and NFW (for M33) profiles with the parameters considered by~\citet{2019PhRvD..99l3027D} as benchmark, while using the rest to estimate the uncertainty connected to the poor knowledge of DM density distribution in M31 and M33. The benchmark profiles are marked with a dagger ($^\dagger$) symbol in Table~\ref{tab:profiles} and \ref{tab:profiles2}. 

\begin{figure*}[!h]
    \centering
    \includegraphics[width=1.1\textwidth]{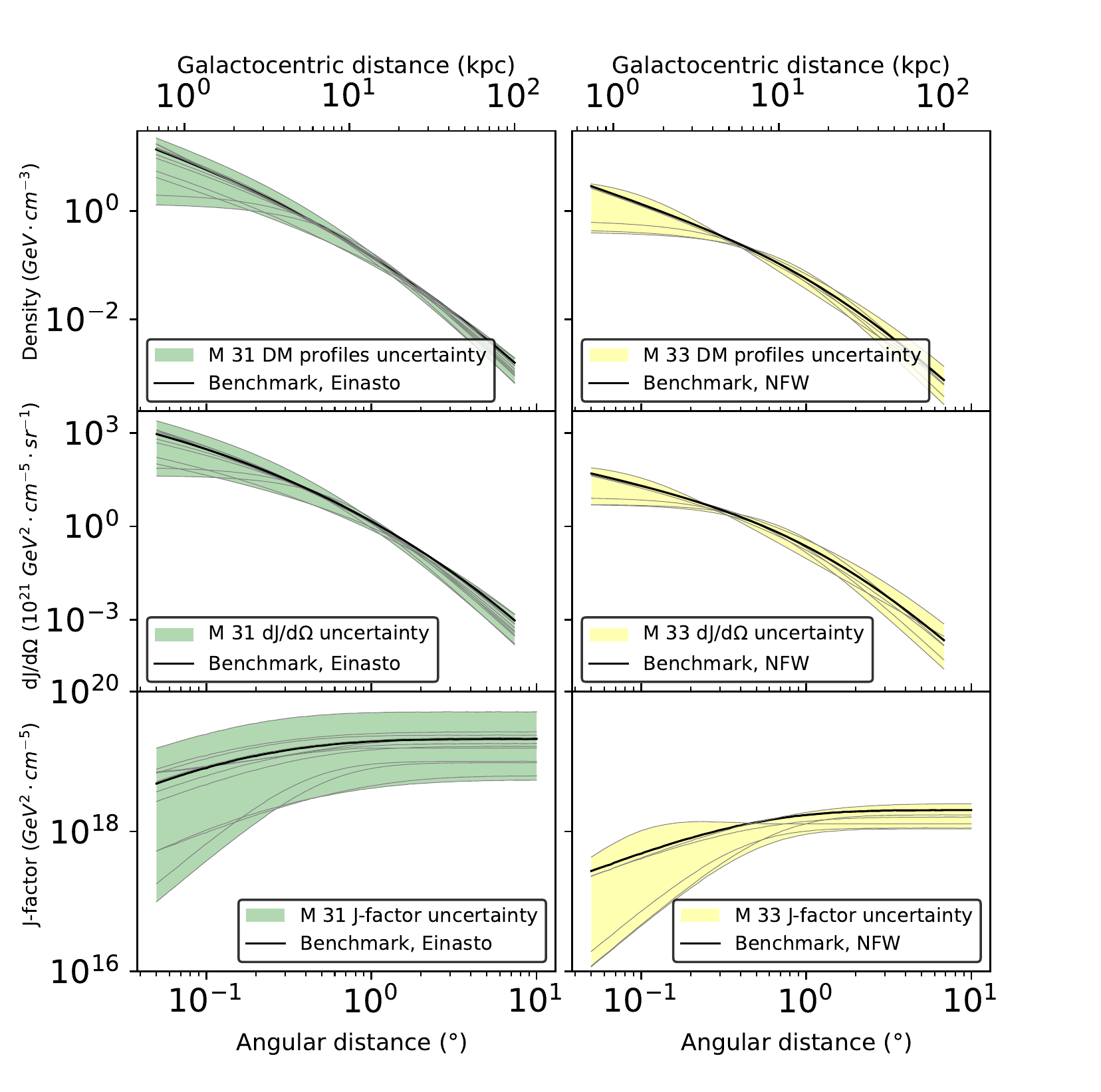}
    \caption {DM density profiles, see Tab.~\ref{tab:profiles} and \ref{tab:profiles2} for the relevant profiles, and corresponding J-factors as a function of the angular distance from the center of the objects of interest (left panels: M31, right panels: M33). Upper panels: DM density distribution profiles. Central panels: Differential J-factor values dJ/d$\mathrm{\Omega}$ of the corresponding DM profiles. Lower panels: Integrated J-factor values over solid angle for the corresponding profiles. The benchmark DM models (Einasto for M31 and NFW for M33, see section~\ref{sec:bench_models}) used in the analysis are highlighted with the bold black solid line in all panels.}
    \label{JJ1}
    
\end{figure*}

\begin{figure*}[!h]
    \centering
    \includegraphics[width=0.43\textwidth]{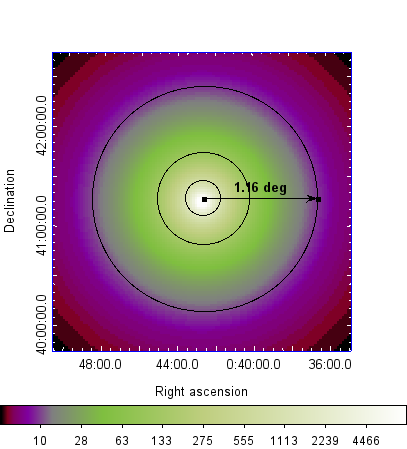}
    \includegraphics[width=0.55\textwidth]{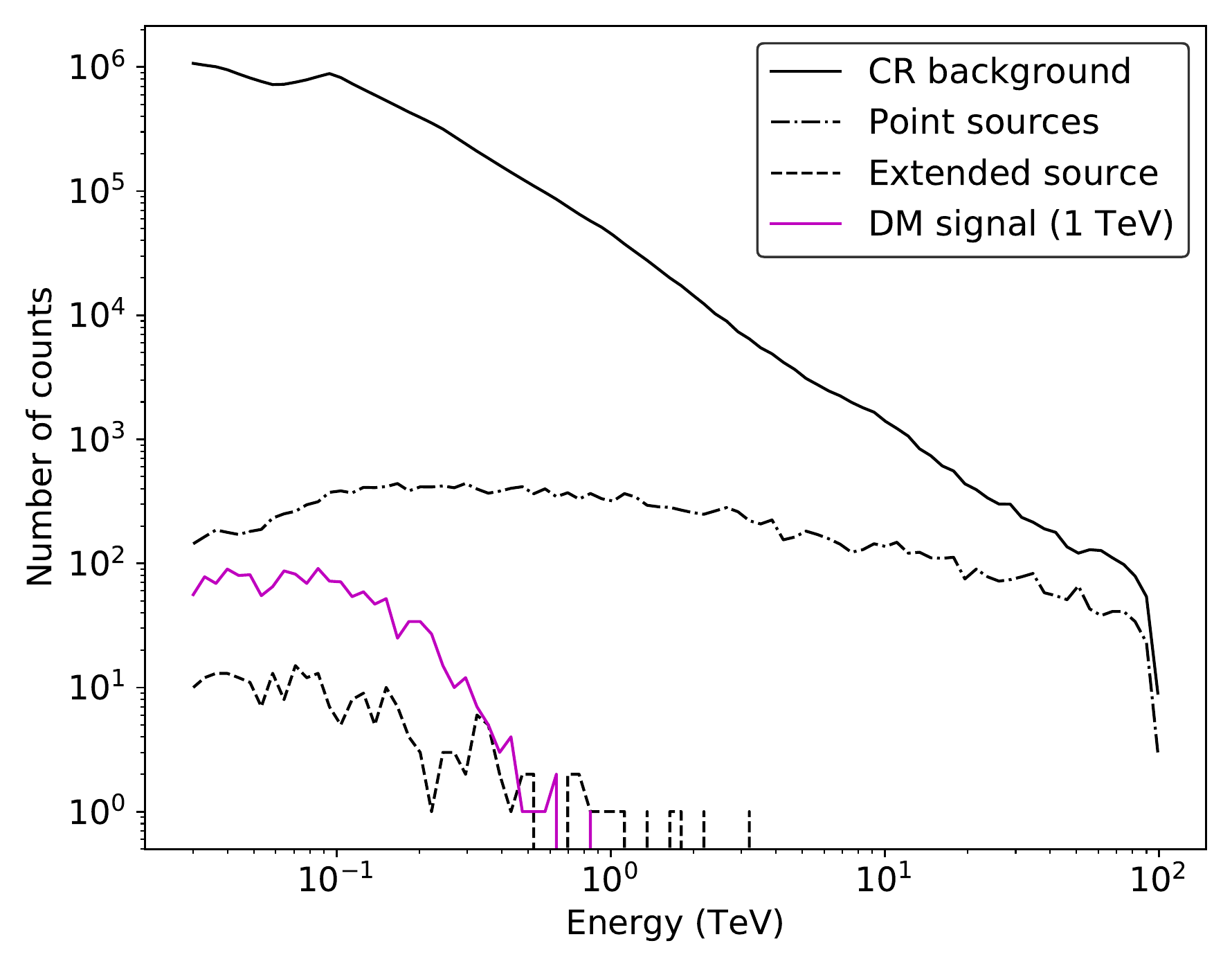}
    \caption[DM source template for M31 benchmark model and number of counts]{Left: DM source template for the Einasto profile that was presented in \citet{2019PhRvD..99l3027D} (benchmark model). The image is centered on M31. The color illustrates $J$-factor in units of~$\mathrm{10^{20}~GeV^{2}\cdot cm^{-5}}$. Black contours present the distances at which $J$-factor decreases by a factor of 10, 100, 1000 in comparison to its maximum.
    Right plot: Expected number of photons predicted by CTA simulations towards M31 direction as a function of energy from the sources contributing to the observed signal for a single realization of the data. An extended source that represents the contribution from M31 bulge with the parameters reported in \citet{2019ICRC...36..570K}. The stacked contribution from 6 point sources present in the FoV of CTA, as shown in the left panel of Fig.~\ref{position}. The DM signal corresponds to the benchmark density model of M31, 1~TeV DM mass, $b\bar{b}$ annihilation channel and $\langle\sigma\upsilon\rangle=2.05\cdot10^{-24}~\mathrm{cm^{3}\cdot s^{-1}}$, corresponding to the value the CTA will be capable of excluding at $95$\% c.l. level, see section~\ref{simple-result})}
    \label{fig:template}
\end{figure*}

All considered density profiles and $J$-factors as functions of distance from the object's center are shown with thin lines in top and bottom panels of Fig.~\ref{JJ1} correspondingly. The benchmark profiles correspond to the thick black line. In order to avoid any underestimation of the actual DM density uncertainty, we calculated the fractional uncertainty $\delta\rho_s/\rho_s=0.04$ of the Einasto profile of M31, which stands for the uncertainty in the determination of the benchmark model itself and appears to be is negligible compared to the green uncertainty region, as shown in the top left panel of Fig.~\ref{fig:template}, which correspond to the actual uncertainty of DM density distribution within the object of interest. The two-dimensional representation of M31 $J$-factor for the reference density profile is presented in the left panel of Fig.~\ref{fig:template}.

\subsection{Gamma-ray emission of conventional astrophysical origin}\label{background}
In addition to a suggested signal from annihilating DM the emission from M31 and M33 directions is complemented by several types of conventional astrophysics fore- and background (point-like and/or diffuse) emissions. These backgrounds include MW galactic diffuse emission, as well as contributions from galactic and extra-galactic sources.

In the particular case of M31, we note also the possible presence of a diffuse GeV/TeV signal originating from the nucleus/bulge/disk of this galaxy. Such a signal was reported in M31 observations in the GeV band with \flat~\citep{2017ApJ...836..208A,2018ApJ...862...79E,2019ICRC...36..570K,2021arXiv210206447A}. The signal is moderately extended (radial disk with a radius of $0.4\degree$ \citep{2017ApJ...836..208A,2019ICRC...36..570K}) and characterised by a relatively soft best-fit spectrum ($2.8\pm0.3$, \citet{2019ICRC...36..570K}). The observed emission can be interpreted within several models, including millisecond pulsar population~\citep{2018ApJ...862...79E} or annihilating/decaying DM particles \citep{2019ICRC...36..570K}.

In the case of M33, no extended signal was clearly observed despite several dedicated searches~\citep{2010JCAP...04..014A,2017ApJ...836..208A,2019ICRC...36..570K,2019PhRvD..99l3027D,2020ApJ...901..158X}. At the same time, a presence of a relatively weak ($6-7\sigma$ significance) source was reported at the position of M33~\citep{m33_fermi}. However, we did not include the above source in this analysis since it is not included in either the 3FHL or 4FGL catalogs. 

\begin{figure}[!h]
    \centering
    \includegraphics[width=0.474\textwidth]{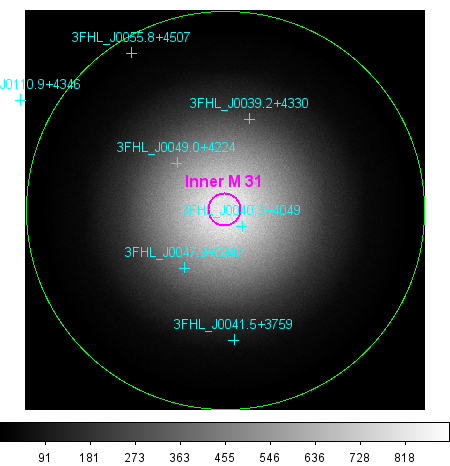} 
    \includegraphics[width=0.495\textwidth]{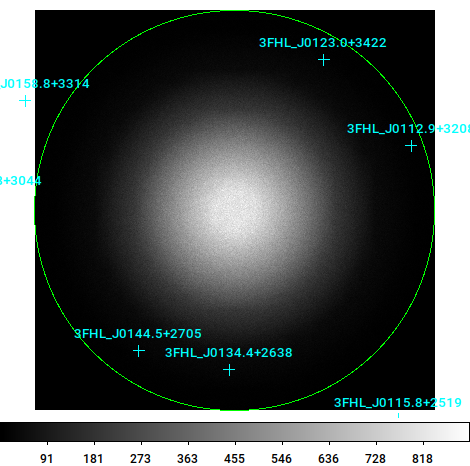}    
    \caption[View of the known gamma-ray sources towards M31 and M33]{The simulated CTA maps of gamma-like events to the direction of M31 (left) and M33 (right) galaxies (background not subtracted). The positions of known \flat sources detected above 10~GeV from 3FHL catalogue are shown with cyan crosses. The green circle illustrates the FoV of CTA with a $5\degree$ radius. The magenta ellipse at the left panel presents the extended Inner M31 source -- a radial disk with $0.4\degree$ radius~\citep{2019ICRC...36..570K}.}
    \label{position}
    
\end{figure}

To estimate the contribution from the galactic and extragalactic sources we consider nearby known GeV sources with the spectrum potentially continuing to the TeV band. The searches within 3FHL~\citep{Ajello_2017} (7-yr \flat catalogue of sources detected above 10~GeV), 4FGL-DR2~\citep{2020ApJS..247...33A} (12 years catalogue of \flat sources detected above 0.1~GeV) and TeVCAT\footnote{\url{http://tevcat2.uchicago.edu/}} resulted in six and four point sources within the CTA FoV ($5^\circ$ radius) around M31's and M33's positions respectively, as shown in Fig~\ref{position}.

The basic parameters of the considered sources (coordinates, shape, spectral parameters -- normalisation and spectral index, possible type, and multiwavelength identification) are summarised in Table~\ref{ALL-SOURCES}.

\section{Data simulation and analysis}\label{III}
\subsection{Data simulation}\label{simulation}
The simulation of the data was performed with \texttt{ctools} v.1.7.3\footnote{\url{http://cta.irap.omp.eu/ctools/}} simulation and analysis package,
in energy band 0.03~TeV -- 100~TeV. For both M31 and M33, we consider 100~hours long observation centered at the corresponding objects. As discussed in section~\ref{zero} the current strategy for the forthcoming CTA DM observations includes 150~hours observation towards M31, while for M33 there is no planning to this point \citep{2019scta.book.....C}. In this work, for simplicity, we consider 100~hours simulation time for both considered targets.

For the simulation and subsequent data analysis, we utilised \texttt{prod3b-v2} instrument response functions (IRF)\footnote{When the analysis was at its latest stages \texttt{prod5.v0.1} IRF were released. We argue that the new IRFs do not affect significantly the derived results, see e.g. left panel of Fig.~\ref{fig:limits-allchannels} for the comparison of \texttt{prod3b-v2} vs. \texttt{prod5.v0.1} results.}. These IRFs are available for North (La Palma) and South (Paranal) CTA sites and a set of zenith angles which additionally determine the proper low-energy threshold $E_{min}$ for the analysis.

The minimal zenith angle under which a source with declination $\delta$ is visible from an observational site with latitude $lat$ is given by $mza = |\delta - lat|$. For the reasonable quality observations, we additionally demand $mza<60^\circ$.
The basic parameters used for the data simulation of M31 and M33 galaxies are summarized in Tab.~\ref{zenithangles}.

\begin{table*}[]
    
    \begin{tabular}{|c|c|c|c|c|c|}
    \hline \hline
        \textbf{Galaxy}&\textbf{Exposure}&\textbf{Culmination} & \textbf{$prod3b-v2$ IRF} &\textbf{$\mathrm{E_{min}}$}\\
         &h& (North/South) &    & TeV\\
        \hline
        \textbf{M31}&100   & $12\degree$/$66\degree$ & $North\_z20\_50h$/ --& 0.06 / --\\
        \hline
        \textbf{M33} &100 & $2\degree$/$56\degree$ & $North\_z20\_50h/South\_z60\_50h$& 0.06/0.13\\
        \hline \hline
    \end{tabular}
    
    \caption{The basic parameters of M31 and M33 used for the CTA data simulation. The first column corresponds to the name of the galaxy/target, while the second one expresses the minimal zenith angle by which each target can be observed by each CTA array. The instrument response functions, based on the minimal zenith angle estimation, used for each target and each array, are reported in the third column. The last column represents the minimum energy, based on the latest CTA suggestions, that one should consider when performing simulation using different IRFs.}
    \label{zenithangles}
\end{table*}

For the simulations of the data, we explicitly consider that the following sources in the FoV of the CTA are contributing to the observed emission:

\begin{itemize}
    \item Residual Cosmic ray background (implemented as ``CTAIrfBackground'' within \texttt{ctools}).
    \item Astrophysical sources in the near vicinity of the target. These include fore/background point-like sources from 3FHL catalogue~\citep{Ajello_2017} of sources detected by  \flat above 10~GeV as well as extended source presenting the extended emission from inner parts of M31 reported by~\citet{2019ICRC...36..570K}. The basic information about all included sources is summarized in Tab.~\ref{ALL-SOURCES}.
\end{itemize}

Given the high galactic latitudes of both galaxies selected for the analysis, we neglected the contribution from the galactic diffuse emission. Aiming in constraining the parameters of WIMP DM  (potentially not present in the real data) we did not include any contribution from the annihilating DM to the simulated data. 

We simulated the data according to the model described above using \texttt{ctobssim} (50 random realizations of the data, defined by initial random seed) and \texttt{ctmodel} (one, non-randomized realization of the model) \texttt{ctools} routines. The data simulated with \texttt{ctmodel} was used as Asimov dataset for the analysis described in detail below.

\subsection{Data analysis}\label{analysis}
We analysed the simulated data within the frame of standard binned CTA data analysis\footnote{See e.g. \href{http://cta.irap.omp.eu/ctools/users/tutorials/quickstart/index.html}{binned analysis tutorial}} implemented in \texttt{ctools}. We additionally cross-checked the results with an alternative implementation of the analysis used by~\citet{2021JCAP...01..057A}.

The analysis relies on the fitting of the 3D (spatial and spectral) model of the region to the data\footnote{See details of the implementation of the fitting procedure at \href{http://cta.irap.omp.eu/ctools/users/user_manual/likelihood.html}{ctools website}.}. The model used for the analysis of simulated data included all components used for data simulation (residual CR background as well as astrophysical sources in the FoV of the CTA). Aiming to constrain the parameters of annihilating WIMP DM we additionally included in the model the template (DM source) presenting such a contribution. DM source template for a set of considered annihilation channels and WIMP masses was composed of spectral and spatial parts as described in Sec.~\ref{signal}. The spectral part is based on approximations of WIMP annihilation spectra by~\citet{Ciafaloni_2011,2011JCAP...03..051C,2012JCAP...10E.001C}. For the spatial part of the model, we considered several DM profiles for each of the considered objects, see Tab.~\ref{tab:profiles} and Tab.~\ref{tab:profiles2}. $J$-factors for each of the considered models were calculated with the publicly available \texttt{CLUMPY}~v.3.0.0 code~\citep{Charbonnier_2012,H_tten_2019}.
The results presented below were obtained with \texttt{ctulimit} task and correspond to $95\%$ confidence level upper limits on $\langle\sigma\upsilon\rangle$. 

To determine the mean expected CTA sensitivity for annihilating DM signal in the considered objects, we utilised the Asimov dataset described in Sec.~\ref{simulation} with benchmark DM profiles for each of the considered annihilation channels ($b\bar{b}, W^{+}W^{-}$~and~$\tau\bar{\tau}$). We also used 50 simulated randomized datasets to estimate the uncertainties connected to the random realizations of the simulated/observed datasets.

We additionally identify several sources of systematic uncertainties which can significantly affect the derived limits. These sources include effects of: \textit{(i)} poor knowledge of DM profiles in selected objects; \textit{(ii)} poor modeling of nearby fore/background astrophysical sources; \textit{(iii)} poor knowledge of CTA response functions (including effective area, PSF and residual CR background mismodeling). We study the contributions from each of these effects in detail and summarize the used approaches and derived results in what below.

\section{Results}\label{IV}
\subsection{Expected CTA sensitivity}\label{simple-result}

In this section we present the main results of our analysis, CTA sensitivity to DM signal from M31 and M33, using Asimov data set and considering the benchmark DM-source models described in section~\ref{sec:target_selection}. Under ''CTA limits'' we mean the limits which CTA could provide for the case of no signal observation, i.e. CTA sensitivity for a detection of annihilating DM signal.

\begin{figure*}[!h]
    \centering
    
    \includegraphics[width=0.5\textwidth]{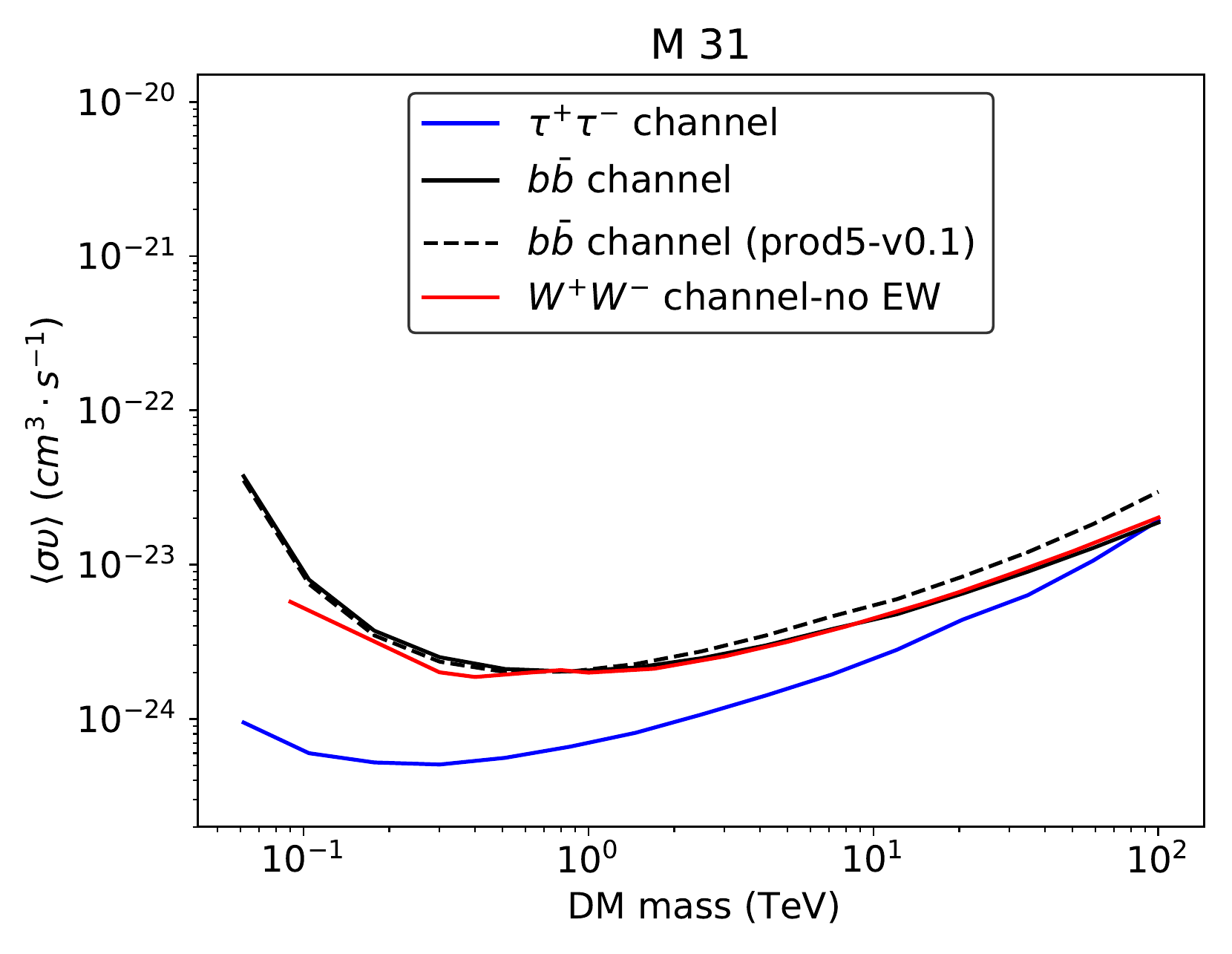}
    \includegraphics[width=0.49\textwidth]{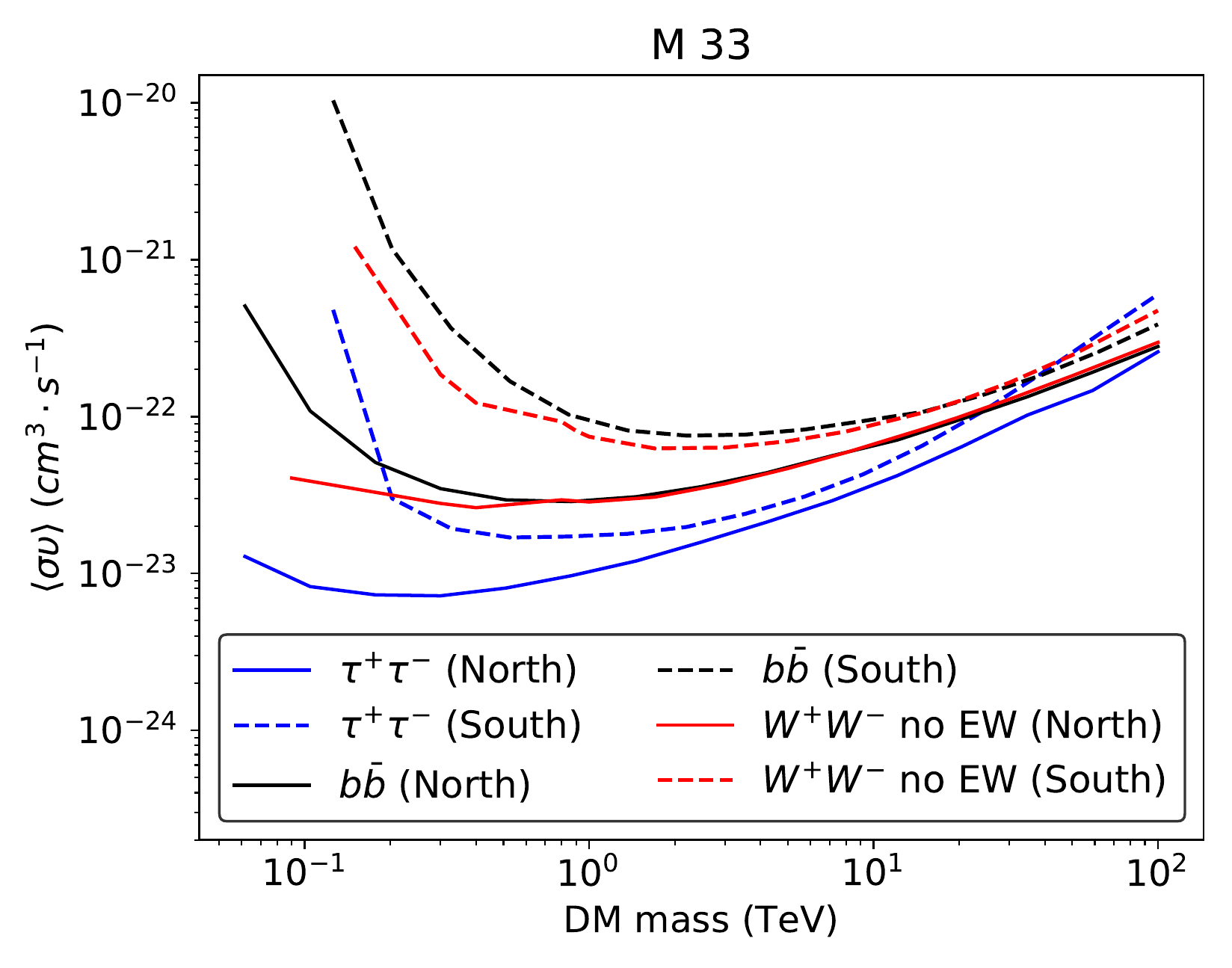}
    \caption{$95\%$ confidence level expected CTA upper limits on velocity-averaged WIMPs annihilation cross-section from 100~h long observations of M31 (left panel) and M33 (right panel) with CTA-North. The benchmark density profiles are assumed in both cases. Blue, black and red curves correspond to $\tau^+\tau^-$, $b\bar{b}$ and $W^+W^-$ annihilation channels. The black dashed line at the left panel illustrates upper limits results for the $b\bar{b}$ annihilation channel when utilizing \texttt{prod5-v0.1} IRFs. Dashed curves at the right panel correspond to the limits which could be obtained with 100~h observations of the same objects with CTA-South array.}
    \label{fig:limits-allchannels}
    
\end{figure*}

Fig.~\ref{fig:limits-allchannels} presents $95\%$ confidence level expected upper limits for the weighted velocity annihilation cross-section for $b\bar{b}$, $\tau^+\tau^-$ and $W^+W^-$ annihilation channels for 100~h long observations of M31 and M33. We note, however, that more constraining results can be obtained when one considers the contribution of DM subhalos, since the presence of such DM substructures can moderately and/or significantly boost the DM annihilation signal, depending on the modeling approach employed. A detailed modeling of such DM substructures in M31 field halo and how the presence of the latest can provide constraints tightening on the corresponding cross-sections is discussed in Appendix~\ref{appendB}.
In what below, we additionally discuss the impact of several considered sources of systematic uncertainties on the derived limits.

\subsection{Effects of uncertainties on the DM density distribution}\label{prof-uncert}

To assess the uncertainties arising from the incomplete knowledge of the DM density distribution in the selected objects, we identified several M31/M33 DM profiles reported in the literature and repeated the analysis described in Sec.~\ref{analysis} for each DM profile. The complete list of the considered profiles is given in Tab.~\ref{tab:profiles} and Tab.~\ref{tab:profiles2}. 
Corresponding $J$-factor profiles as functions of the distance to the center of the corresponding galaxy are shown in Fig.~\ref{JJ1}. As demonstrated by this figure, the difference in $J$-factors' profiles at some distances can reach an order of magnitude, resulting in about the same potential worsening of the derived limits on the WIMP annihilation cross-section. 

We conclude that the current measurements of the DM density distribution in M31 and M33 carry sizable uncertainties, especially so in the central regions of these galaxies. These DM-density uncertainties are one of the dominant systematic ones which can substantially worsen any derived results. An additional source of $J$-factor uncertainty is the contribution from the DM annihilating in the MW halo. Our estimations show that this contribution at positions of M31 and M33 galaxies is sub-dominant in comparison to the DM-annihilation signal in M~31 and M~33, see Appendix~\ref{appendC}. Correspondingly, in what below, we neglect the contribution from DM annihilating in the MW halo.

\begin{figure*}[!h]
    \centering
    \includegraphics[width=0.497\textwidth]{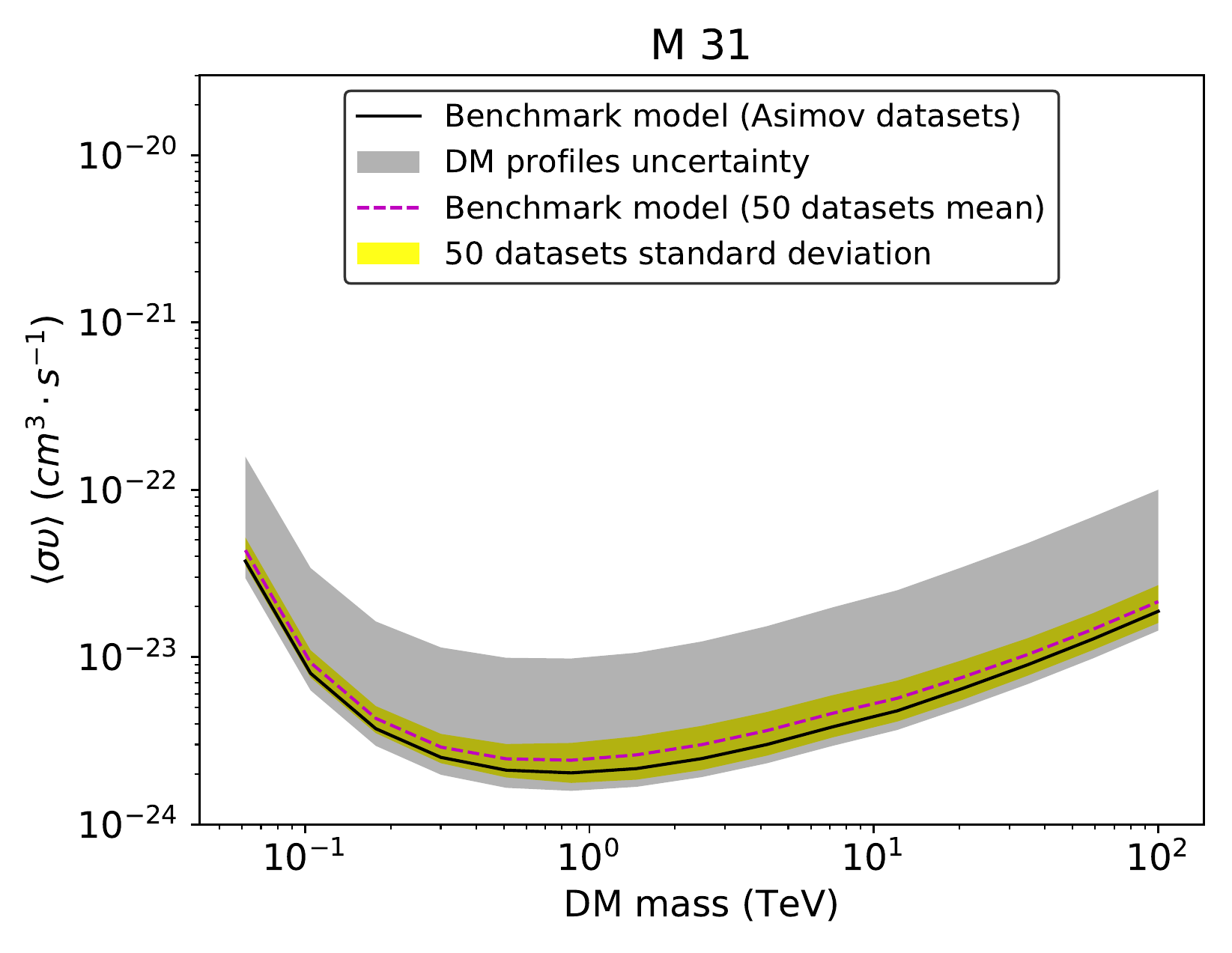}
    \includegraphics[width=0.495\textwidth]{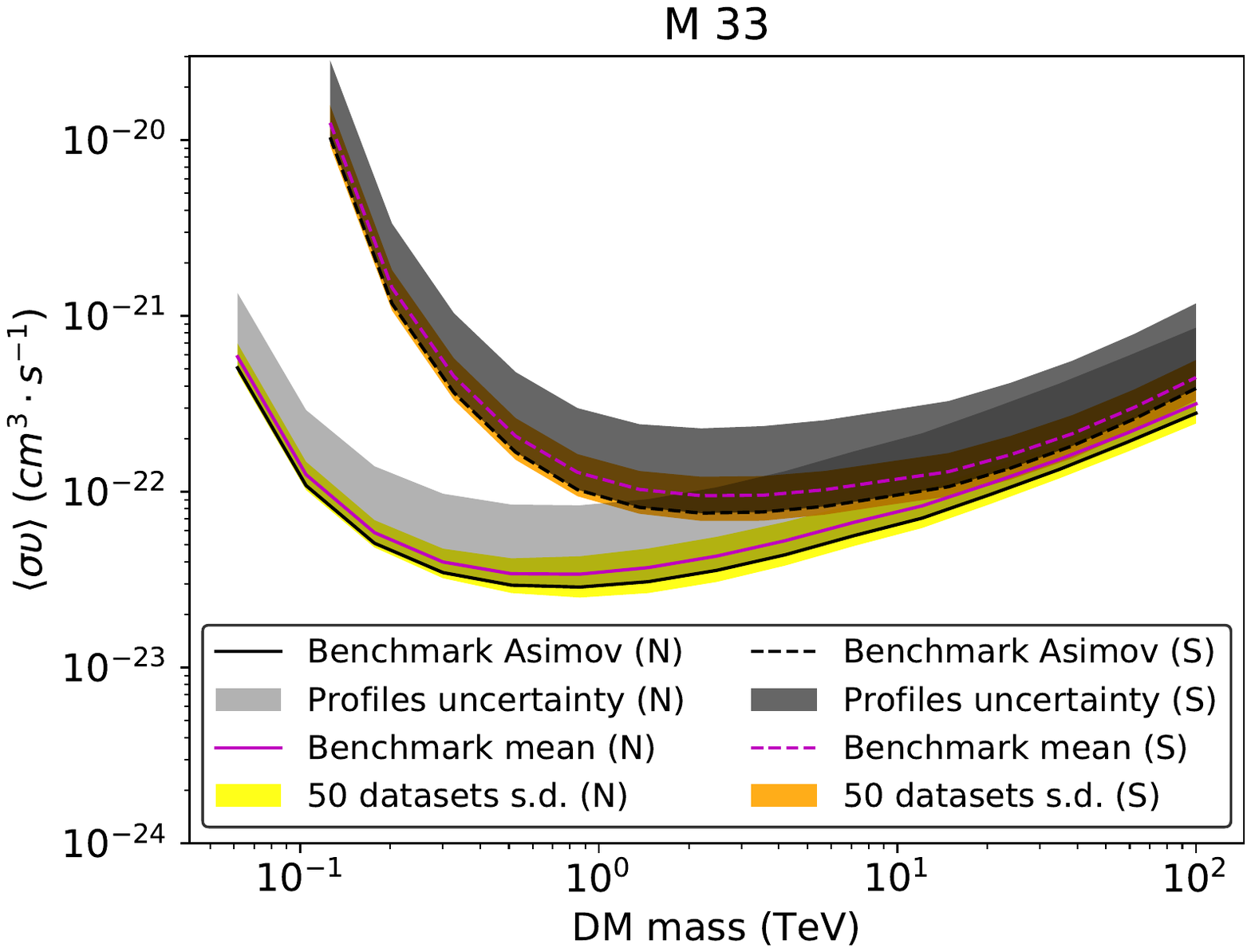}
    \caption{Left: The grey shaded region resent the range of limits on velocity-averaged annihilation cross-section ($b\bar{b}$ annihilation channel) for 100~h long M31 (left panel) observations with CTA-North for the set of DM density profiles summarized in Tab~\ref{tab:profiles} and \ref{tab:profiles2}. The black solid lines correspond to expected limits for the benchmark models of DM density profiles, based on the simulated Asimov data set. The magenta line and yellow shaded regions correspond to the mean values and standard deviations of the expected limits from 50 random statistically independent realizations of the data. Right panel: same for M33. Results for the CTA-North and CTA-South arrays are indicated with (N) and (S) correspondingly.}
    \label{BB}
    
\end{figure*}

The results of our analysis are shown in Fig.~\ref{BB} for the benchmark $b\bar{b}$ channel and in Fig.~\ref{TT} and \ref{WW} for $\tau^+\tau^-$, and $W^+W^-$ channels correspondingly. The shaded regions correspond to the envelope of the upper limits obtained for all considered profiles.
For the illustration in Fig.~\ref{BB} we additionally present the uncertainty region connected to random data realizations. These regions are shown with yellow and orange colors for the Northern and Southern CTA site respectively.

\subsection{Effects of uncertainties on the astrophysical backgrounds}\label{astro-uncert}

Additional uncertainty during the analysis can arise from the presence of poorly modelled point-like or diffuse sources in the CTA's FoV. In the case of M31, we note the presence of a central diffuse astrophysical source, \citep[see e.g.][]{2019ICRC...36..570K} which potentially could mimic DM annihilation signal and spoil CTA sensitivity for DM studies in this object.

To assess the impact of the presence of the discussed point-like/diffuse sources, we performed simulations/analysis of the data similar to Sec.~\ref{analysis} with and without explicit modelling of the sources (see Tab.~\ref{ALL-SOURCES}). The obtained in both cases upper limits coincide with a discrepancy of 10~\%. This allows us to conclude on the relative unimportance of the contribution of nearby sources for the presented results.

\subsection{GeV emission from the Inner M31 bulge}\label{InnerM31}

The GeV \flat spectrum energy distribution of the central astrophysical source in M31 is shown in Fig.~\ref{InnerM31} with light-grey (reported by \citet{2017ApJ...836..208A}) and orange points (reported in the recent study \citet{2021arXiv210206447A}). The red line and shaded region show the best-fit power law parameters of~\citet{2019ICRC...36..570K} above 1~GeV. Assuming that the M31 spectrum continues to the TeV band as a power law, our modelling shows that the CTA will not be able to detect this source. Blue upper limits present 95\% c.l. flux upper limits that CTA could reach with a 100~h long observation of the region. For the illustration with a green line, we show the spectrum of 12.1~TeV DM annihilating to $b\bar{b}$ channel. The strength of the signal corresponds to the 95\% upper limit reported in Fig.~\ref{BB} for annihilation cross-section at this mass. 

We additionally explore the possibility of a break/hardening of M31 spectral index at $\sim 5$~GeV energies, as marginally indicated by \flat spectral points. Fig.~\ref{InnerM31}, right panel, shows the TS of the detection of M31 central source as a function of the assumed spectral index. We conclude that the CTA will be able to detect M31 with $TS\gtrsim 9$ only if its spectrum is harder than $\sim 2.4$, while high significance detection ($TS\gtrsim 25$) is achieved only if its spectrum is harder than $\sim 2.3$. The corresponding power law for high-significance detection is shown in the left panel with a black solid line.

\begin{figure*}[!h]
    \centering
    \includegraphics[width=0.51\textwidth]{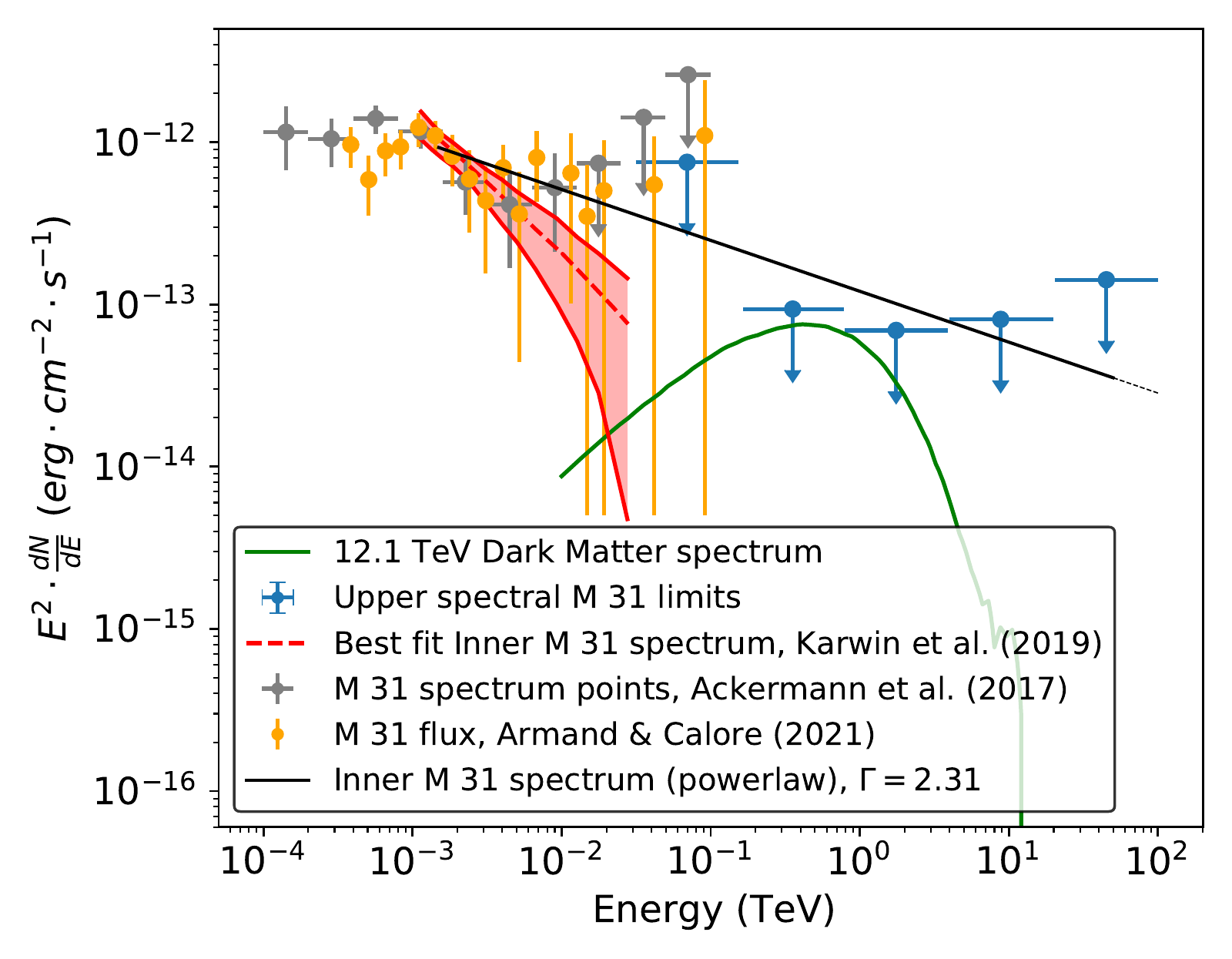}
    \includegraphics[width=0.48\textwidth]{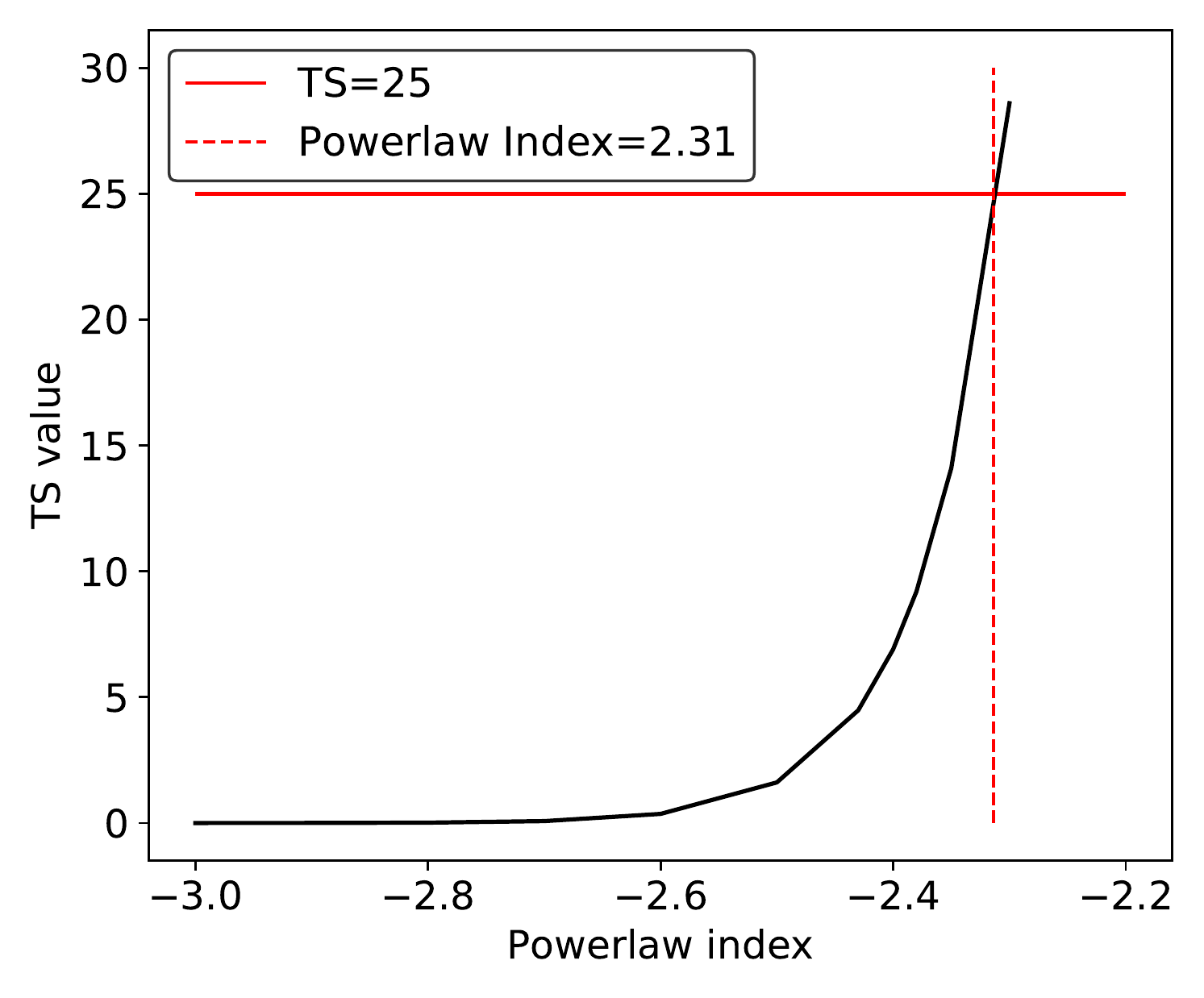}
    \caption[Spectrum and upper limits spectrum of M31]{Left: \flat spectrum and CTA sensitivity to the inner M31 source. The red dashed line inside the red butterfly diagram corresponds to the best-fit powerlaw model for an analysis in the energy range of $10^{-3}$ to $10^{-2}~\mathrm{TeV}$ of the Inner M31 component of M31 galaxy (spectrum: powerlaw, spatial model: radial disk of $0.4\degree$ radius) reported in \citet{2019ICRC...36..570K}. Grey points show the results of \citet{2018ApJ...862...79E},\citet{2017ApJ...836..208A}. The orange points correspond to M31 spectrum with disk-like M31 model~\citep{2021arXiv210206447A}. The blue upper limits present the upper spectral limits on Inner M31 emission (M31 bulge) that CTA could provide with a 100~h-long observation of M31 region. 
     The green solid curve present the annihilation spectrum of 12.1~TeV WIMP DM (benchmark M31 density profile, $b\bar{b}$ channel, $\langle\sigma\upsilon\rangle=4.78\cdot 10^{-24}~\mathrm{cm^{3}\cdot s^{-1}}$).
     The black solid line highlights the case in which M31 could be detected at $5\sigma$ significance level. Normalization of the line matches one reported by~\citet{2019ICRC...36..570K} at 1.5~GeV and continues to higher energies as a powerlaw with the slope $2.31$.\newline
     Right: Detection test-statistics value for M31 assuming that its spectrum matches one reported by~\citet{2019ICRC...36..570K} at 1.5~GeV and continues to higher energies as a powerlaw with the given slope. }
    \label{InnerM31}
    
\end{figure*}

\subsection{Impact of systematic uncertainty}\label{syst-uncert}
\label{sec:syst}
In this section, we discuss the impact of the systematic uncertainties of the instrumental origin and/or related to miss-identified CR on the derived results.
Aiming this, we adopt two different approaches to describe systematic uncertainties in the modeling of the data. Ιn general, systematic uncertainties arise from imperfectly known or poorly controlled instrument characteristics. E.g., the energy-dependent under(over)-estimation of the effective area uncontrollably changing with time can induce artificial spectral features and consequently lead to the false-positive detection of annihilating-DM signal.

In both methodologies, briefly summarized below, we assume 0.1 ($10\%$) energy scale and $0.1^\circ$ spatial scale systematic uncertainties of CTA. These values are close to the characteristic ones for currently operating facilities, such as H.E.S.S.~\citep{2006A&A...457..899A}.
We however explore also lower levels of systematics -- 1\% and 3\% to illustrate the gain of the decreased level of systematics which can be achieved with the next-generation instruments. 

\paragraph{Systematics via likelihood function modification.}
The contribution of the systematic uncertainties can be accounted for by multiplying the predicted signal by scale parameters and profiling their likelihood over the value of the selected parameters. We select scale parameter $\boldsymbol{\alpha}=\alpha_{ij}$ for which we assume Gaussian nuisance likelihood with an i,j-independent variance $\sigma_{\alpha}^{2}$. In principle, the distributions are considered log-normal fainting to zero as a $\alpha$ goes to zero. Based on that scale parameter, we utilize for our analysis the following modified likelihood function \citep[see e.g., ][]{2015JCAP...03..055S}

\begin{equation}
    \boldsymbol{L(\mu,\alpha\mid n)}=\prod_{i,j}\frac{(\mu_{ij}\alpha_{ij})^{n_{ij}}}{\sqrt{2\pi}\sigma_{\alpha}n_{ij}!}e^{-\mu_{ij}\alpha_{ij}}e^{-\frac{(1-\alpha_{ij})^{2}}{2\sigma_{a}^{2}}}
    \label{likelihood2}
\end{equation}

\begin{figure*}[!h]
    \centering
    \includegraphics[width=0.8\textwidth]{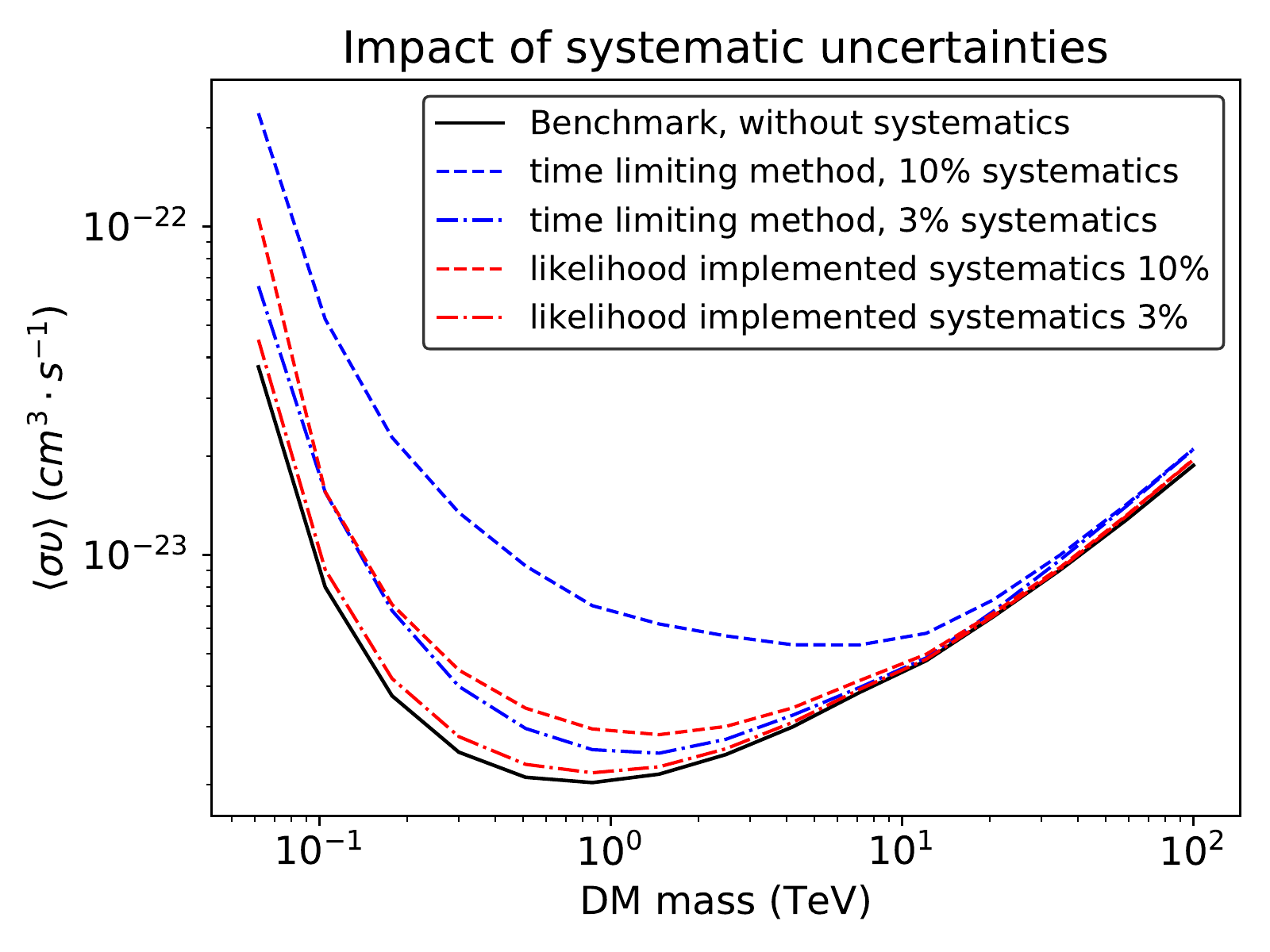}
    \caption[M31 upper limits for the benchmark model in the presence of systematic uncertainties] {M31 upper limits comparison for the benchmark model with and without the presence of systematic uncertainties ($b\bar{b}$ channel). The black solid line corresponds to the upper limits without the presence of systematic uncertainties. The blue/red dashed lines correspond to the upper limits in the presence of systematics ($10\%$ energy scale, $0.1\degree$ spatial scale) following the time-constraining methodology and the likelihood implemented systematics method respectively. The blue/red dash-dotted lines have the same representation but for $3\%$ energy scale.}
    \label{SYST-RES}
    
\end{figure*}

Such a modification of the likelihood provides the opportunity for upper limit derivations when systematic uncertainties (e.g., effective area) enter linearly the calculation of the total signal. The obtained upper limits are presented with the red dashed and dash-dotted line in Fig~\ref{SYST-RES}, for $10\%$ and $3\%$ systematic uncertainty respectively.\\
\paragraph{Systematics via exposure constraining.}
Alternatively, one can address the impact of systematic uncertainties by modeling them via limiting the statistic of the data. 
The observations of the same constant in time phenomena for a time period $t$ result in relative statistical errors scaling $\propto t^{-1/2}$. E.g., for a source with a constant with time count rate $r$~cts/s the observed after time $t$ number of photons would be $N=rt$ with corresponding relative statistical uncertainty $dN_{stat}/N=N^{-1/2}\propto t^{-1/2}$ decreasing with increasing of observational time. We define the relative systematic uncertainty $\alpha$ as $dN_{syst}/N=\alpha$ which remains constant and does not decrease with the increase of observational exposure. This type of uncertainty can reflect poorly controllable behavior of the instrument, e.g. energy dependent quasi-random variations of the effective area during the observation. 

To treat the systematic uncertainty we propose to limit the observational time to the characteristic value for which $dN_{syst}=dN_{stat}$, i.e., to stop the observation as soon as expected systematic uncertainty becomes equal to the statistical one. Longer observation will lead only to the decrease of statistical uncertainty, which will become sub-dominant in comparison to systematical one. 

The requirement $dN_{stat} = dN_{syst}$ can be reformulated in terms of the maximal number of observed photons as $N_{max}=\alpha^{-2}$. We note that $N_{max}$ should not be treated as the total number of photons received during the observation, but rather as a number of photons in the smallest possible statistically independent energy/spatial bins. We note also that generally speaking, the level of systematic $\alpha = \alpha(E)$ can be a function of energy.

The spatial $\delta\theta$ and energy $\delta E$ resolutions of the instrument naturally define statistically independent energy and spatial bins. To properly identify the time for which at a given energy $dN_{stat}=dN_{syst}$ we propose that the observation at energies $[E; E+\delta E]$ should be stopped as soon as the number of photons in any spatial bin of size $\delta\theta$ reached $\alpha^{-2}(E)$. This allows to have in each of the statistically independent spatial/energy bins the number of photons dominated by statistics uncertainty and thus neglect the presence of systematics.

The characteristic values of CTA energy and spatial resolutions are $\sim 10$\% and $\sim 0.1^\circ$ correspondingly. Accordingly, we split 0.03~TeV to 100~TeV simulation energy range over a number of energy and spatial bins. We explicitly limit the observing time when systematic uncertainty becomes equal to the statistical one. I.e. at energies $[E; 1.1\cdot E]$ we stop the observation as soon as any spatial bin of $0.1^\circ$-radius accommodates more than $N=100$~photons for $\alpha=0.1$ and $N=1111$~photons for $\alpha=0.03$. 
We note, that for most observational cases, the highest number of photons in spatial bins at any energy is reached in the spatial bin centered at the center of FoV of the CTA. Due to off-axis decrease of CTA effective area, this position is characterised by the strongest level of the residual cosmic-ray background. In the absence of bright astrophysical sources in the FoV, this background is obviously dominating the observed emission. 

\begin{figure*}[!h]
    \centering
    \includegraphics[width=0.497\textwidth]{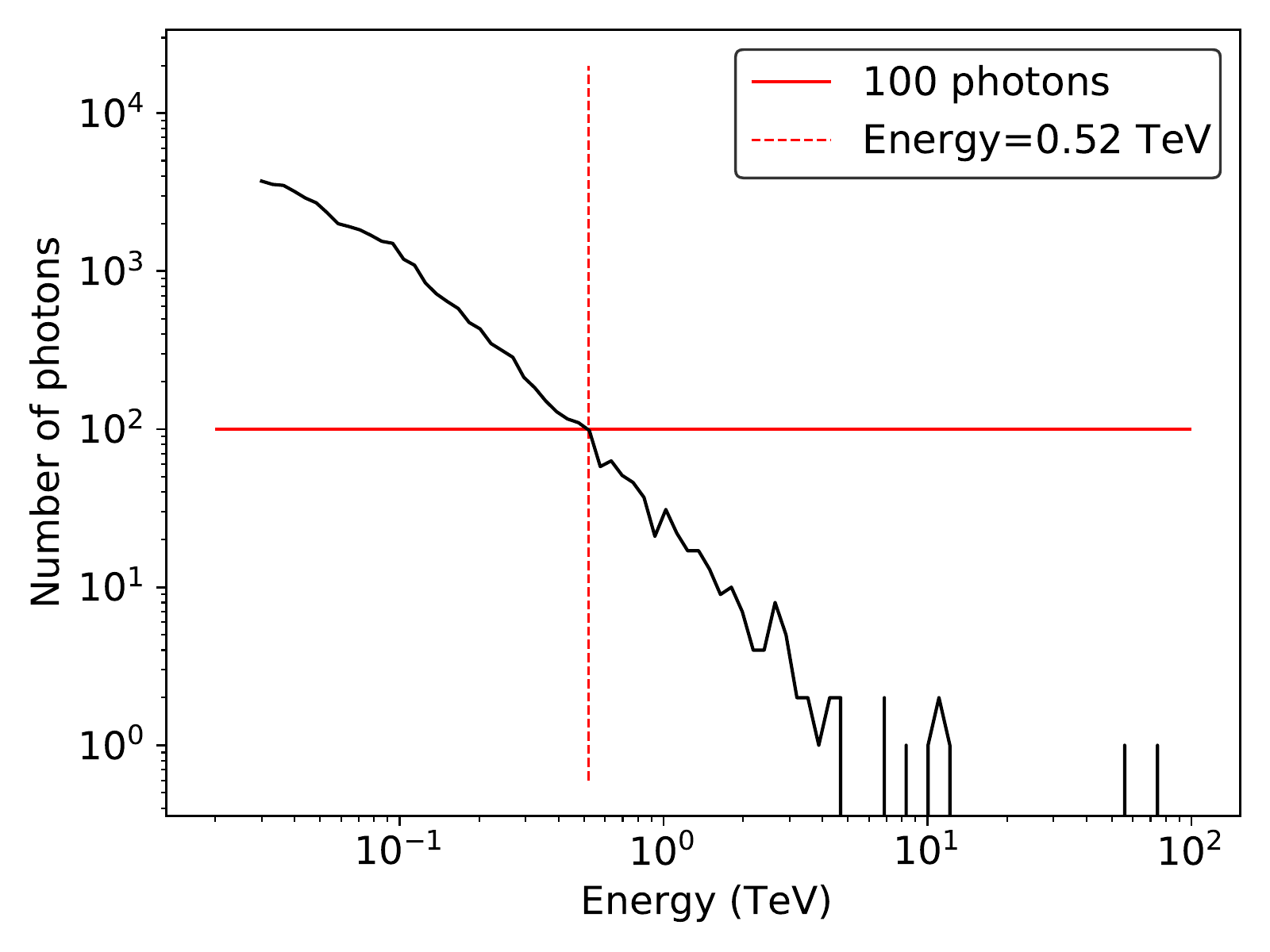}
    \includegraphics[width=0.495\textwidth]{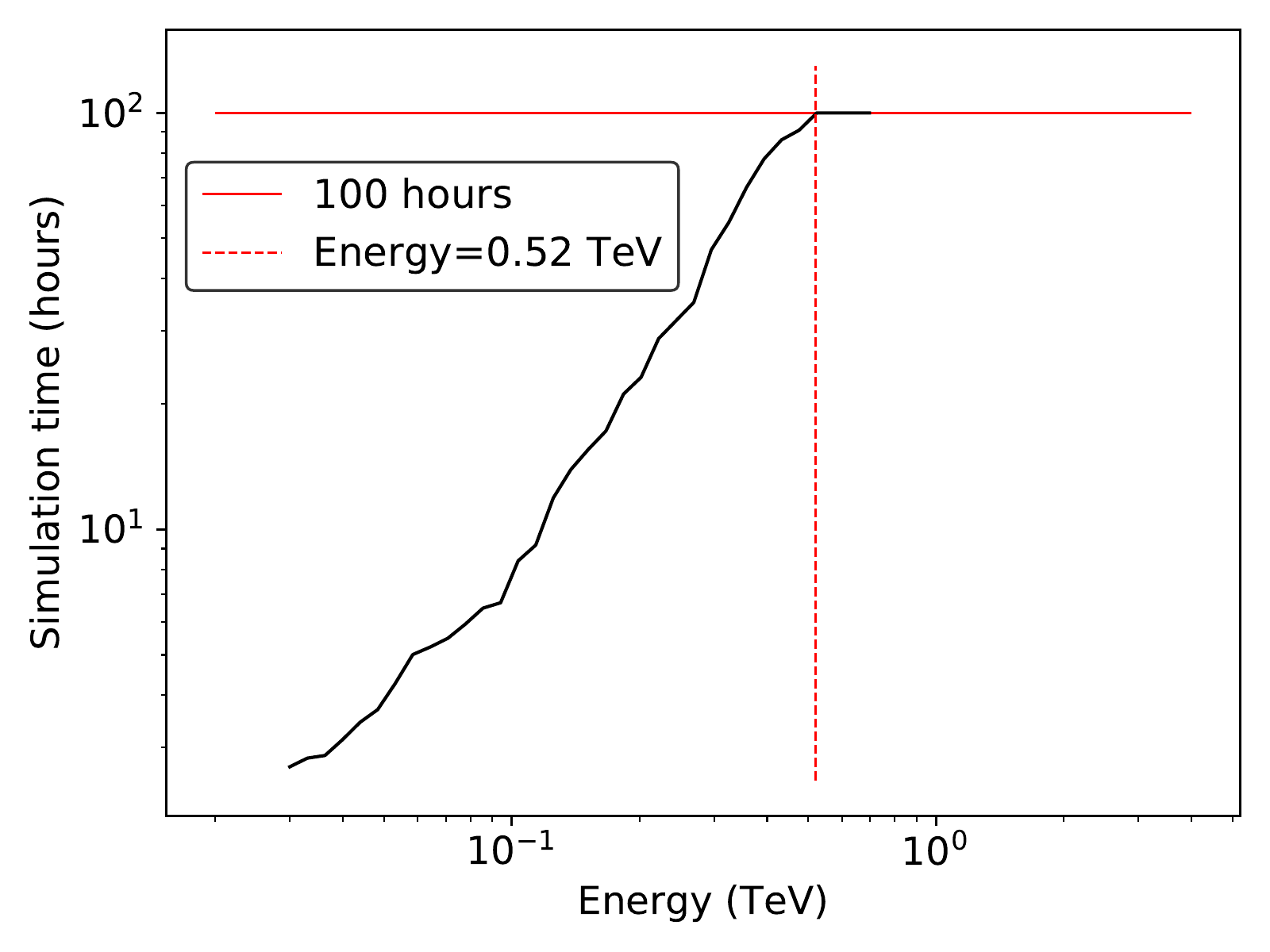}
    \caption[Methodology for systematic uncertainties: Number of photons and simulation time as a function of the energy] {Left: Number of detected photons in the 10\% energy width bins in 0.03 -- 100~TeV energy band. The solid red line shows 100 photons, the characteristic value for which systematic and statistical uncertainties are equal for 10\% systematic uncertainty ($\alpha=0.1$). The energies $>0.52$~TeV(red dashed vertical line) are dominated by statistical uncertainties.\newline Right: Simulation time in the considered energy bins required for systematic uncertainty to be equal or smaller than the statistical one. The solid red line corresponds to 100~hours of simulation time. }
    \label{NU-TI}
    
\end{figure*}

We show the number of photons as a function of energy, for a region of $0.1\degree$ spatial scale centered at the center of CTA FoV in Fig.~\ref{NU-TI}, left panel. The red horizontal line illustrates $100$~photons -- the characteristic number of photons at which the observation should be stopped for the systematic level $\alpha=0.1$. The right panel of the figure presents the exposure required to reach 100~photons in the considered bin as a function of energy. This illustrates that the considered level of systematic affects only the low-energy part of the CTA data. Namely, any energy/spatial bin ($10$\% energy width and $0.1^\circ$ spatial scale) at above $\gtrsim 0.5$~TeV for 100~h long observation does not accommodate more than 100~photons. Correspondingly, at these energies, we performed standard binned analysis (assuming bin size $\sim 10$\% energy width and $\sim 0.1^\circ$ spatial scale) for 100~h Asimov dataset.

At lower energies, we performed dedicated, time-limited simulations of Asimov datasets for each of the considered 10\% energy width bins in a way similar to the simulations described above. For each of the considered bins, including the above-threshold bin at $E>0.5$~TeV we performed the standard binned analysis and build log-likelihood profiles as a function of DM template normalization (proportional to $\langle\sigma\upsilon\rangle$). Adding log-likelihood profiles for all energy bins, we built a joint log-likelihood profile which allowed us to constrain DM annihilation cross-section as a function of DM mass for such energy-dependent exposure observation. 

The results of this approach are summarised in Fig~\ref{SYST-RES}, for $10\%$ and $3\%$ systematics respectively (blue dashed and dot-dashed lines correspondingly). The red lines present the results of the systematic treatment based on the modification of the log-likelihood function described above for similar values of systematic uncertainties. Despite the entirely different approaches considered and generally different treatment of the systematics, we found the results to be broadly consistent with each other at lower levels of systematics ($1\%$ or $3\%$). We note, that the results are not totally identical since the two distinct strategies suggest different origin of the systematic uncertainty and treat it differently. E.g. the first discussed strategy is based on the modification of the log-likelihood function which applies to the whole energy range of the analysis whereas the second one has its basis on the exposure time constraining which only affects the lower energy regimes where the systematic uncertainty dominates over the statistical uncertainty. We note that the Night Sky Brightness maps (NSB) indicate a higher level of emission in the direction of M31, and thus the enhanced background at the location of the Galaxy indicates an even higher level of systematic uncertainty.

We would like to note also that the considered ``systematics via exposure constraining'' approach allows us to identify the scale of systematics which does not affect the observations at 100~h timescale. Namely, the maximum number of the photons in $0.1^\circ$, 10\% energy-width bins seen in simulation is $\lesssim 10^4$, which translates to the systematic level of $\sim 1$\%. We argue that for the lower values of systematic, the 100~h observations will not be affected by considered systematics.

The discussed approach to the systematic treatment allows also to identify the most effective sharing of the observational time between different instruments of the CTA array. Consisting of large (LST), medium (MST), and small (SST) size telescopes CTA observatory can perform observations by its different sub-arrays sensitive to low (LST), intermediate (MST), and high (SST) energies. The exposure time required to reach a given number of photons per energy/spatial bin is typically an increasing function of energy, see e.g. Fig.~\ref{NU-TI}, right panel. For a given level of systematics, this allows to vary the observational time, making it the shortest for the LST and longest for SST telescopes without loss of the scientific outcome of observation. The freed telescopes' time can be used for observations of other targets.

\section{Conclusions and Discussion}\label{V}
Along with the MW, M31, and M33 are the two largest spiral galaxies in the Local group. The proximity of these two galaxies permits detailed studies of DM distribution, showing the DM dominated nature of these objects and making them interesting targets for annihilating WIMPs searches with current and upcoming observational facilities.

In this work, we performed 100--hour long simulation of these galaxies with the next-generation TeV observatory CTA aiming to study the prospects of detecting annihilating DM within these objects. Where applicable under ''CTA limits/constraints'' we meant the limits which CTA could provide for the case of no signal observation, i.e. CTA sensitivity for a detection of annihilating DM signal. We report the expected prospects of detection for DM velocity-averaged annihilation cross-section for a set of annihilation channels ($b\bar{b}$, $\tau^+\tau^-$ and $W^+W^-$).
We have paid special attention to the factors that can affect the CTA sensitivity to the expected signal. In particular, we analysed uncertainties connected to \textit{(i)}: the possible astrophysical background emission in the FoV of CTA; \textit{(ii)} the uncertainties of DM density distribution; \textit{(iii)}: imperfect or poor knowledge of the instrument itself, i.e. systematic uncertainties.

We found that the uncertainties on the DM profiles result in the highest uncertainty in the derived prospects. Namely, for the density profiles summarized in Tab~\ref{tab:profiles} the corresponding uncertainty can reach an order of magnitude for certain annihilation channels, see Fig~\ref{BB}, \ref{TT}, and \ref{WW}. We, therefore, argue that the detailed studies of DM distribution in M31 and M33 are essential for the accurate estimate of WIMP annihilation detection within these objects.

Fig.~\ref{fig:limits-allchannels} summarizes the $95$\% constraints derived for the benchmark density profiles for all considered channels for both discussed galaxies. The figure illustrates that the observation of M31 from the Northern (La Palma) CTA site generally provides better constraints in comparison to M33 observations. The best limits are derived for $\tau^+\tau^-$ annihilation channel and reached the level of $5\cdot10^{-25}~\mathrm{cm^{3}~s^{-1}}$ at energies $\sim 0.3$~TeV.

\begin{figure*}[!h]
    \centering
    \includegraphics[width=0.497\textwidth]{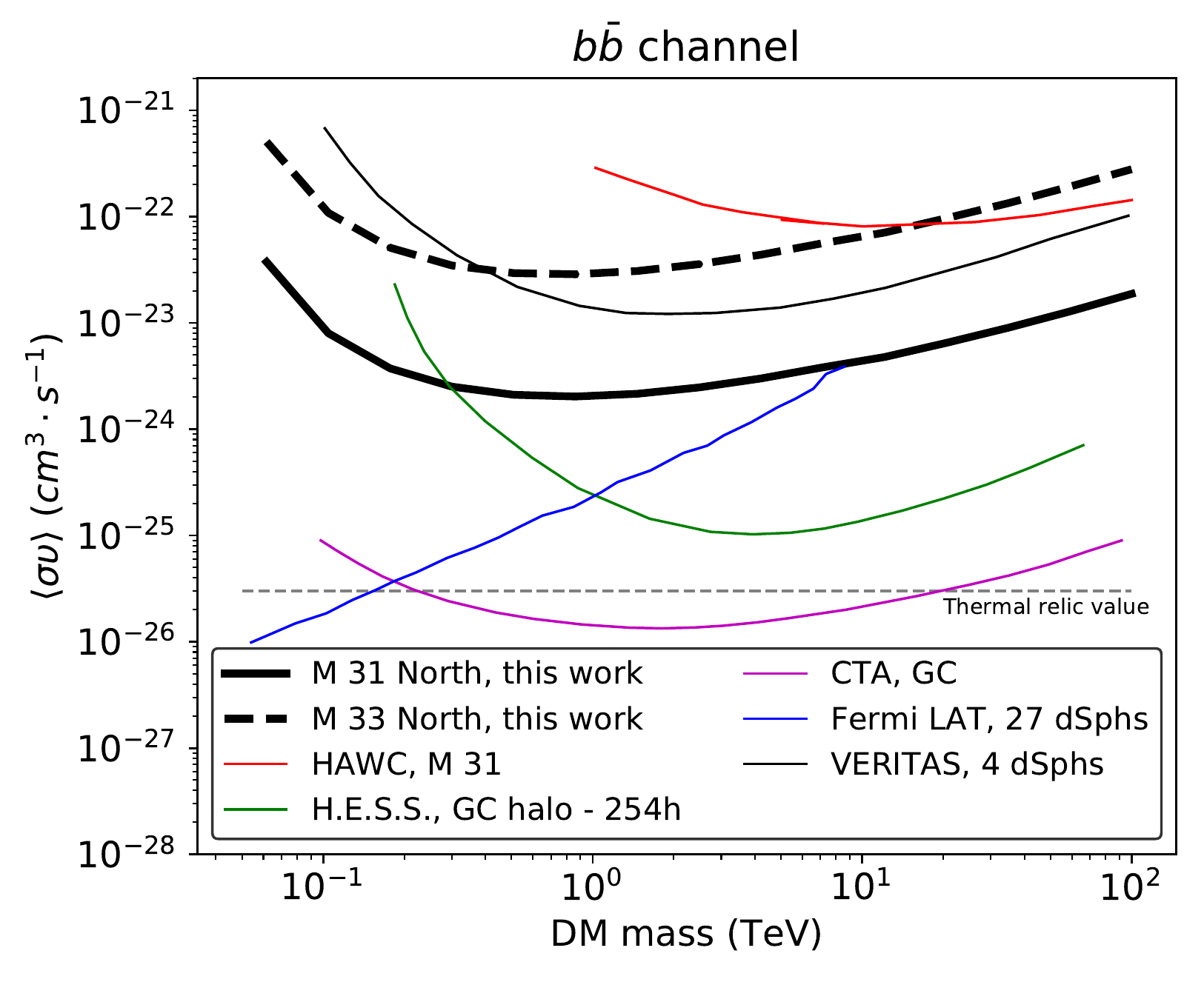}
    \includegraphics[width=0.495\textwidth]{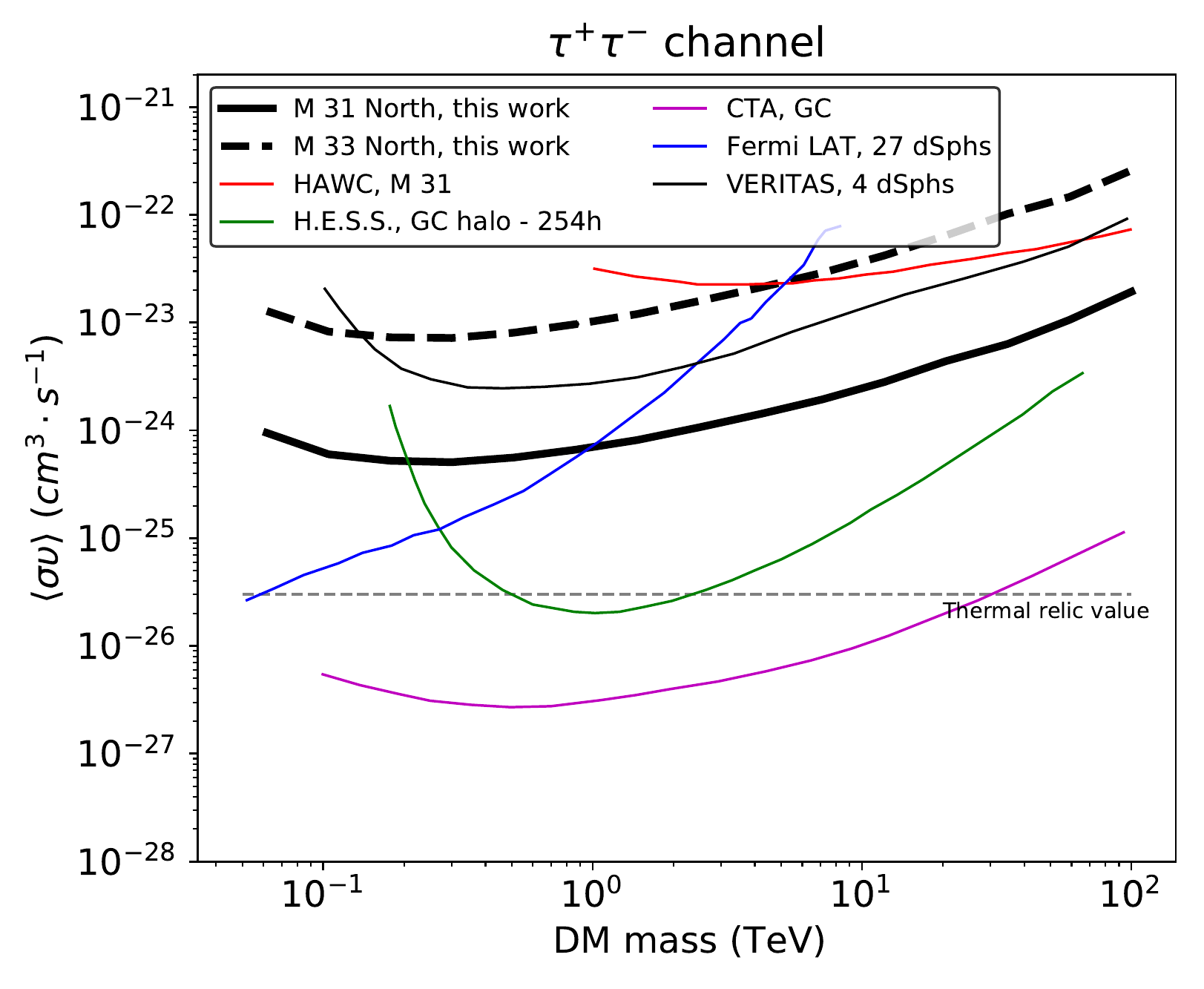}
    \caption{Comparison of constraints on the $b\bar{b}$ (left panel) and $\tau^{+}\tau^{-}$ (right panel) channel of the upper limits of this work -- M31 (solid black line-benchmark model) and M33 (dashed black line-benchmark model) -- with the previous published HAWC limits (red solid line for the same Einasto model that we consider as benchmark model for M31~\citep{hawc_m31}), H.E.S.S. limits of observations of the GC (green solid line~\citep{Abdallah_2016}), the limits of the observations of 27 dSphs of the MW by Fermi satellite (blue solid line~\citep{fermi_dsphs}), the limits from the GC by CTA (magenta solid line~\citep{2019PhRvD..99l3027D}), and the combined analysis of observations of 4 dSphs by VERITAS (cyan solid line~\citep{Archambault_2017}).}
    \label{sumi}
    
\end{figure*}

The 100~h long CTA observations of M31/M33 could improve --by an order of magnitude-- the limits derived by modern facilities from non-observation of the annihilation signal from M31 by HAWC~\citep{hawc_m31} or from 4 dSphs by VERITAS~\citep{Archambault_2017}, see~Fig~\ref{sumi}. At the same time, the observations of the Galactic Center with modern observatories~\citep[see e.g.][]{Abdallah_2016} or with CTA~\citep{2019PhRvD..99l3027D} could be able to provide better constraints. At below 10~TeV energies, the expected limits are also substantially worse than the limits derived from \flat observations of 27~dSphs~\citep{fermi_dsphs}. We argue, however, that taking into account the possible effects of highly uncertain astrophysical background and DM density distribution, the observations of the proposed in this work targets could provide important constraints on WIMP DM parameter space.

The possible astrophysical emitting sources that are reported in the literature for both M31 and M33 do not appear bright enough to strongly affect the derived results. A particular set of simulations and fitting was dedicated to ascertaining whether or not the extended Inner M31 astrophysical source is detectable by CTA. The simple powerlaw fit that we performed resulted in a power index value of 2.3 or harder, for such detection to be possible (see Fig~\ref{InnerM31}).

Finally, the last cause of sensitivity loss, studied in this work, is the systematic uncertainties case. For the characteristic values of the systematic uncertainty of 3--10\% expected for CTA, we compare the results of two approaches of the systematics treatment. One of the approaches is based on the modification of log-likelihood function used for the fitting of the model to the data while the second is based on the constraint of the observational time so the statistical uncertainty becomes comparable to the systematics level, see~\ref{sec:syst}. Although the results of the two different approaches are not identical, this is understandable due to the distinctive nature of the strategies employed. Both methods result in a somewhat comparable sensitivity loss (mainly for lower levels of systematics i.e., $1\%$ and $3\%$) in comparison to no systematic case, see Fig.~\ref{SYST-RES}. The loss affects mostly low DM masses, the limits for which are strongly dominated by low-energy data, where the systematics plays the most significant role. In the case of 10\% systematics the expected loss of sensitivity can reach a factor of 3 indicating potential substantial worsening of CTA limits at low DM masses.

The exposure limiting approach to systematic treatment allows also to identify the systematics level $\alpha_{min}\sim 1$\% at which 100~h long CTA observations will not be sensitive to the systematic effects. Finally, we propose the energy-dependent observational strategy, which allows efficient use of different telescopes from the CTA array in the presence of systematics. Namely, we argue that in this case, the observational time can be selected to be shortest for LST telescopes and the longest for SST ones without compromising the scientific outcome of the observation. The freed telescopes' time can be used for the observation of other targets.

\acknowledgments

We would like to thank Torsten Bringmann for helpful discussions. This work was conducted in the context of the CTA Dark Matter Exotic Physics Working Group. This paper has gone through internal review by the CTA Consortium.
This work has been supported by the Fermi Research Alliance, LLC under Contract No. DE-AC02-07CH11359 with the U.S. Department of Energy, Office of High Energy Physics.
The work of CE is further supported by the ``Agence Nationale de la Recherche'' through grant ANR-19-CE31-0005-01 (PI: F.~Calore).  AS is supported  by the Kavli Institute for Cosmological Physics at the University of Chicago through an endowment from the Kavli Foundation and its founder Fred Kavli. 
AB is supported by the European Research Council (ERC) Advanced Grant ``NuBSM'' (694896). DM is supported by DFG through the grant MA 7807/2-1 and DLR through the grant 50OR2104. The authors acknowledge support by the state of Baden-W\"urttemberg through~bwHPC. 

We gratefully acknowledge financial support from the following agencies and organizations: State Committee of Science of Armenia, Armenia;
The Australian Research Council, Astronomy Australia Ltd, The University of Adelaide, Australian National University, Monash University, The University of New South Wales, The University of Sydney, Western Sydney University, Australia; Federal Ministry of Education, Science and Research, and Innsbruck University, Austria;
Conselho Nacional de Desenvolvimento Cient\'{\i}fico e Tecnol\'{o}gico (CNPq), Funda\c{c}\~{a}o de Amparo \`{a} Pesquisa do Estado do Rio de Janeiro (FAPERJ), Funda\c{c}\~{a}o de Amparo \`{a} Pesquisa do Estado de S\~{a}o Paulo (FAPESP), Funda\c{c}\~{a}o de Apoio \`{a} Ci\^encia, Tecnologia e Inova\c{c}\~{a}o do Paran\'a - Funda\c{c}\~{a}o Arauc\'aria, Ministry of Science, Technology, Innovations and Communications (MCTIC), Brasil;
Ministry of Education and Science, National RI Roadmap Project DO1-153/28.08.2018, Bulgaria; The Natural Sciences and Engineering Research Council of Canada and the Canadian Space Agency, Canada; CONICYT-Chile grants CATA AFB 170002, ANID PIA/APOYO AFB 180002, ACT 1406, FONDECYT-Chile grants, 1161463, 1170171, 1190886, 1171421, 1170345, 1201582, Gemini-ANID 32180007, Chile, W.M. gratefully acknowledges support by the ANID BASAL projects ACE210002 and FB210003, and FONDECYT 11190853; Croatian Science Foundation, Rudjer Boskovic Institute, University of Osijek, University of Rijeka, University of Split, Faculty of Electrical Engineering, Mechanical Engineering and Naval Architecture, University of Zagreb, Faculty of Electrical Engineering and Computing, Croatia;
Ministry of Education, Youth and Sports, MEYS  LM2015046, LM2018105, LTT17006, EU/MEYS CZ.02.1.01/0.0/0.0/16\_013/0001403, CZ.02.1.01/0.0/0.0/18\_046/0016007 and CZ.02.1.01/0.0/0.0/16\_019/0000754, Czech Republic; Academy of Finland (grant nr.317636 and 320045), Finland;
Ministry of Higher Education and Research, CNRS-INSU and CNRS-IN2P3, CEA-Irfu, ANR, Regional Council Ile de France, Labex ENIGMASS, OCEVU, OSUG2020 and P2IO, France; The German Ministry for Education and Research (BMBF), the Max Planck Society, the German Research Foundation (DFG, with Collaborative Research Centres 876 \& 1491), and the Helmholtz Association, Germany; Department of Atomic Energy, Department of Science and Technology, India; Istituto Nazionale di Astrofisica (INAF), Istituto Nazionale di Fisica Nucleare (INFN), MIUR, Istituto Nazionale di Astrofisica (INAF-OABRERA) Grant Fondazione Cariplo/Regione Lombardia ID 2014-1980/RST\_ERC, Italy; ICRR, University of Tokyo, JSPS, MEXT, Japan; Netherlands Research School for Astronomy (NOVA), Netherlands Organization for Scientific Research (NWO), Netherlands; University of Oslo, Norway; Ministry of Science and Higher Education, DIR/WK/2017/12, the National Centre for Research and Development and the National Science Centre, UMO-2016/22/M/ST9/00583, Poland; Slovenian Research Agency, grants P1-0031, P1-0385, I0-0033, J1-9146, J1-1700, N1-0111, and the Young Researcher program, Slovenia; South African Department of Science and Technology and National Research Foundation through the South African Gamma-Ray Astronomy Programme, South Africa; The Spanish groups acknowledge the Spanish Ministry of Science and Innovation and the Spanish Research State Agency (AEI) through the government budget lines PGE2021/28.06.000X.411.01, PGE2022/28.06.000X.411.01 and PGE2022/28.06.000X.711.04, and grants PID2022-139117NB-C44, PID2019-104114RB-C31,  PID2019-107847RB-C44, PID2019-104114RB-C32, PID2019-105510GB-C31, PID2019-104114RB-C33, PID2019-107847RB-C41, PID2019-107847RB-C43, PID2019-107847RB-C42, PID2019-107988GB-C22, PID2021-124581OB-I00, PID2021-125331NB-I00; the ''Centro de Excelencia Severo Ochoa'' program through grants no. CEX2019-000920-S, CEX2020-001007-S, CEX2021-001131-S; the ''Unidad de Excelencia Mar\'ia de Maeztu'' program through grants no. CEX2019-000918-M, CEX2020-001058-M; the ''Ram\'on y Cajal'' program through grants RYC2021-032552-I, RYC2021-032991-I, RYC2020-028639-I and RYC-2017-22665; the ''Juan de la Cierva-Incorporaci\'on'' program through grants no. IJC2018-037195-I, IJC2019-040315-I. They also acknowledge the ''Atracción de Talento'' program of Comunidad de Madrid through grant no. 2019-T2/TIC-12900; the project ''Tecnologi\'as avanzadas para la exploracio\'n del universo y sus componentes'' (PR47/21 TAU), funded by Comunidad de Madrid, by the Recovery, Transformation and Resilience Plan from the Spanish State, and by NextGenerationEU from the European Union through the Recovery and Resilience Facility; the La Caixa Banking Foundation, grant no. LCF/BQ/PI21/11830030; the ''Programa Operativo'' FEDER 2014-2020, Consejer\'ia de Econom\'ia y Conocimiento de la Junta de Andaluc\'ia (Ref. 1257737), PAIDI 2020 (Ref. P18-FR-1580) and Universidad de Ja\'en; ''Programa Operativo de Crecimiento Inteligente'' FEDER 2014-2020 (Ref. ESFRI-2017-IAC-12), Ministerio de Ciencia e Innovaci\'on, 15\% co-financed by Consejer\'ia de Econom\'ia, Industria, Comercio y Conocimiento del Gobierno de Canarias; the ''CERCA'' program and the grant 2021SGR00426, both funded by the Generalitat de Catalunya; and the European Union's Horizon 2020 GA:824064 and NextGenerationEU (PRTR-C17.I1); Swedish Research Council, Royal Physiographic Society of Lund, Royal Swedish Academy of Sciences, The Swedish National Infrastructure for Computing (SNIC) at Lunarc (Lund), Sweden; State Secretariat for Education, Research and Innovation (SERI) and Swiss National Science Foundation (SNSF), Switzerland; Durham University, Leverhulme Trust, Liverpool University, University of Leicester, University of Oxford, Royal Society, Science and Technology Facilities Council, UK;
U.S. National Science Foundation, U.S. Department of Energy, Argonne National Laboratory, Barnard College, University of California, University of Chicago, Columbia University, Georgia Institute of Technology, Institute for Nuclear and Particle Astrophysics (INPAC-MRPI program), Iowa State University, the Smithsonian Institution, V.V.D. is funded by NSF grant AST-1911061, Washington University McDonnell Center for the Space Sciences, The University of Wisconsin and the Wisconsin Alumni Research Foundation, USA.

The research leading to these results has received funding from the European Union's Seventh Framework Programme (FP7/2007-2013) under grant agreements No~262053 and No~317446.
This project is receiving funding from the European Union's Horizon 2020 research and innovation programs under agreement No~676134.

\newpage
\begin{appendices}

\section{Summary of DM profiles and upper limit results for $\tau^{+}\tau^{-}$ and $W^{+}W^{-}$ annihilation channels}\label{appendC}

In this section, we summarize a large sample of DM density profiles reported in the literature for M31 and M33 galaxies. In Tab~\ref{tab:profiles} and \ref{tab:profiles2} we present the basic information on these objects (coordinates, distance, possible CTA observational site) as well as parameters of DM density profiles (scale density and radius) used in this work to estimate the uncertainties connected to density uncertainties in these objects.

\begin{table}[]
     \begin{adjustwidth}{-2.65cm}{}
    \begin{tabular}{|c|c|c|c|c|c|c|c|c|}
    \hline \hline
        \textbf{Galaxy} & \textbf{\thead{l,b\\(~$\degree$)}}  & \textbf{Distance} & \textbf{\thead{CTA \\site}} &\textbf{Profile} & \textbf{$r_s$}  & \textbf{$\rho_s$} & \textbf{$log_{10}[J(0.5\degree)]$}&\textbf{references} \\
        & &  \textbf{kpc} && &\textbf{kpc}  & \textbf{GeV/cm$^3$} & \textbf{GeV/cm$^5$}& \\
        \hline
        \multirow{17}{*}{M31}& \multirow{17}{*}{\thead{121.17,\\ -21.57}}&\multirow{17}{*}{778}& \multirow{17}{*}{North}
        &NFW & $8.18$ & $1.43423\cdot10^{0}$ &19.33 & \cite{geehan06} \\
        &&&&NFW & $12.5$ & $6.60504\cdot10^{-1}$ & 19.16&\cite{2007arXiv0707.4374T} \\
        &&&&NFW & $34.6\pm2$ & $8.46\cdot10^{-2}$ & 18.52 &\cite{2015PASJ...67...75S} \\
        &&&&NFW & $16.5\pm1.5$ & $4.18\cdot10^{-1}$ & 19.09 &\cite{tamm12} \\
        &&&&NFW & $30.2^{12.1}_{-8.8}$ & -  & -&\cite{2014ApJ...789...62H} \\
        &&&&NFW (M31a) & $12.94$ & - &- & \cite{2005ApJ...631..838W} \\
        &&&&NFW (M31b) & $14.03$ & - &- & \cite{2005ApJ...631..838W} \\
        &&&&NFW (M31d) & $17.46$ & - & -& \cite{2005ApJ...631..838W} \\
        &&&&NFW & $7.63$ & $2.342132\cdot10^{0}$ & 19.67 &\cite{2014PASJ...66L..10K} \\
        \cline{5-8}
        &&&&Burkert & $9.06\pm0.53$ & $1.4\cdot10^{0}$ & 18.71&\cite{tamm12} \\
        &&&&Burkert & $6.86$ & $2.171312\cdot10^{0}$ & 18.83 &\cite{2007arXiv0707.4374T} \\
        \cline{5-8}
        &&&&Einasto$^\dagger$ & $178\pm18$ & $3.08\cdot10^{-4}$ & 19.24&\thead{\cite{tamm12}/\\\cite{2019PhRvD..99l3027D}} \\
        &&&&Einasto & $387\pm44$ & $5.32\cdot10^{-5}$ & 18.51 &\cite{tamm12} \\
        &&&&Einasto & $135.0$ & $5.1246\cdot10^{-4}$ &  19.36&\cite{2007arXiv0707.4374T} \\
        \cline{5-8}
        &&&&Moore & $31.0\pm3$ & $5.54\cdot10^{-2}$ & 19.19&\cite{tamm12} \\
        &&&&Moore & $25.0$ & $7.7818\cdot10^{-2}$ &  19.15&\cite{2007arXiv0707.4374T} \\
        \cline{5-8}
        &&&&SIS & $>8.1$ & - & - &\cite{2014ApJ...789...62H} \\
        \cline{5-8}
        &&&&HYB & $>117.5$ & - &  -&\cite{2014ApJ...789...62H} \\
        \hline \hline
    \end{tabular}
    \end{adjustwidth}
    \caption{A summary of basic parameters of M31. The table summarizes Galactic coordinates of M31 (l,b), the distance to the object, visibility from Northern (La Palma) or Southern (Chile) CTA site as well as parameters of DM density distribution (profile type, characteristic densities $\rho_s$ and radii $r_s$, and the J-factor log-posterior assuming integration over a circular region with angular radius of $0.5\degree$). The benchmark profile is highlighted with a dagger ($^\dagger$) symbol, see text for the details.}
    \label{tab:profiles}
\end{table}

\begin{table}[]
      \begin{adjustwidth}{-2.1cm}{}
    \begin{tabular}{|c|c|c|c|c|c|c|c|c|}
    \hline \hline
        \textbf{Galaxy} & \textbf{\thead{l,b\\(~$\degree$)}}  & \textbf{Distance} & \textbf{\thead{CTA \\site}} &\textbf{Profile} & \textbf{$r_s$}  & \textbf{$\rho_s$} & \textbf{$log_{10}[J(0.5\degree)]$}&\textbf{references} \\
        & &  \textbf{kpc} && &\textbf{kpc}  & \textbf{GeV/cm$^3$} & \textbf{GeV/cm$^5$} & \\
        \hline
        \multirow{7}{*}{M33}& \multirow{7}{*}{\thead{133.61,\\ -31.33}}&\multirow{7}{*}{840}& \multirow{7}{*}{North/South}
        & NFW &$35$ & $5.74\cdot10^{-2}$ & 18.14 &\cite{2010ApJ...709L..32B} \\
        & &&& NFW$^\dagger$ &$22.41$ & $0.1\cdot10^{0}$ &  18.13&\cite{2019PhRvD..99l3027D} \\
        &&& & NFW &$20.78$ & $0.1\cdot10^{0}$ &  18.05&\cite{2017MNRAS.468..147L} \\\cline{5-8}
        &  &&& Burkert  &$12$ & $4.2\cdot10^{-1}$ &  17.86&\cite{2010ApJ...709L..32B} \\
        & &&& Burkert &$7.5$ & $6.83\cdot10^{-1}$ & 17.87 &\cite{2017MNRAS.468..147L} \\
        & &&& Burkert &$9.6$ & $4.669\cdot10^{-1}$ & 17.17 &\cite{2019PhRvD..99l3027D} \\\cline{5-8}
        &&&& Pseudo-Iso &$1.39$ & $4.04\cdot10^{0}$ & 18.11 &\cite{2011ISRAA2011E...4S} \\
        \hline \hline
    \end{tabular}
    \end{adjustwidth}
    \caption{A summary of basic parameters of M33. The table summarizes Galactic coordinates of M31 (l,b), the distance to the object, visibility from Northern (La Palma) or Southern (Chile) CTA site as well as parameters of DM density distribution (profile type, characteristic densities $\rho_s$ and radii $r_s$, and the J-factor log-posterior assuming integration over a circular region with angular radius of $0.5\degree$). The benchmark profile is highlighted with a dagger ($^\dagger$) symbol, see text for the details.}
    \label{tab:profiles2}
\end{table}

\begin{figure*}[!h]
    \centering
    \includegraphics[width=0.497\textwidth]{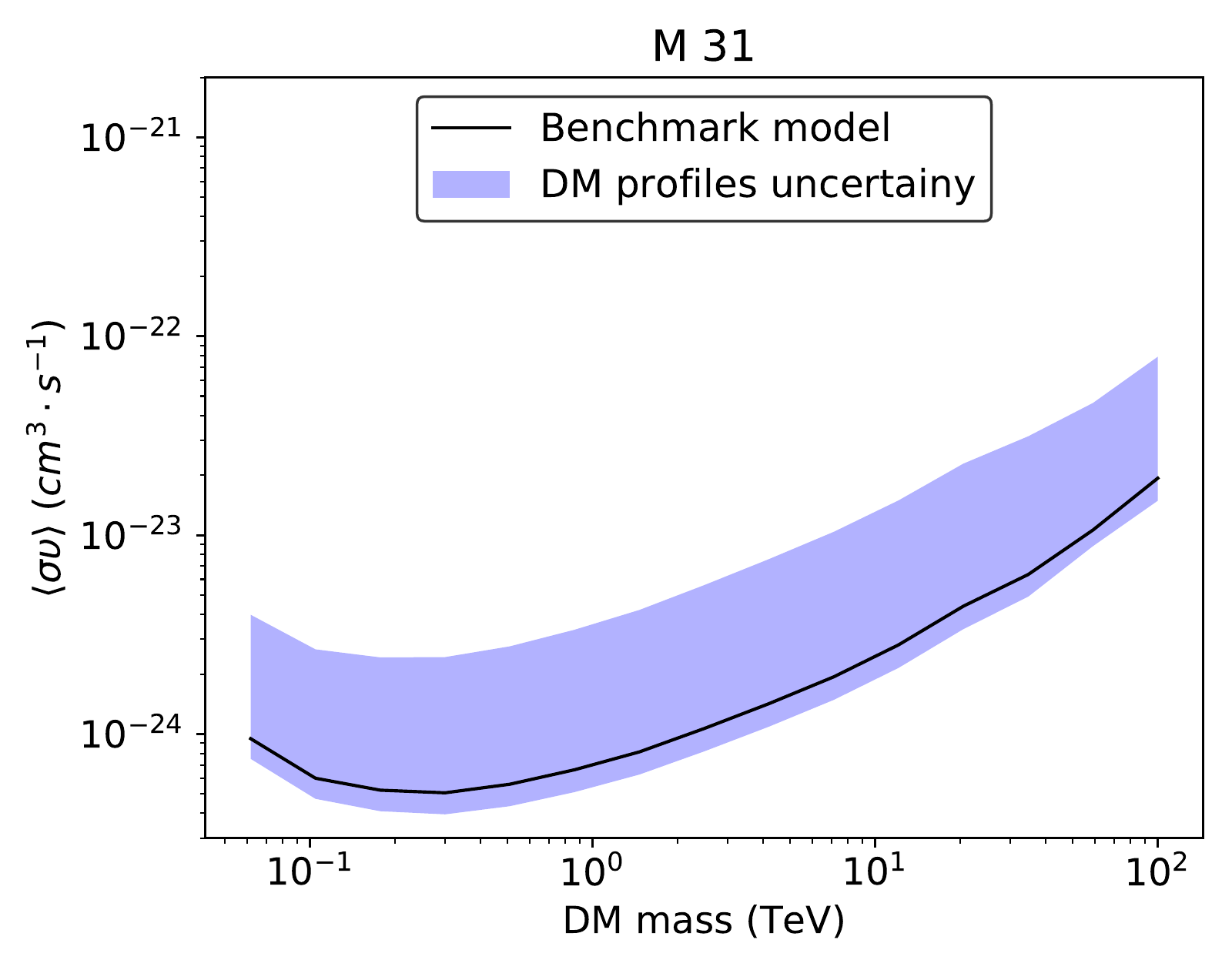}
    \includegraphics[width=0.495\textwidth]{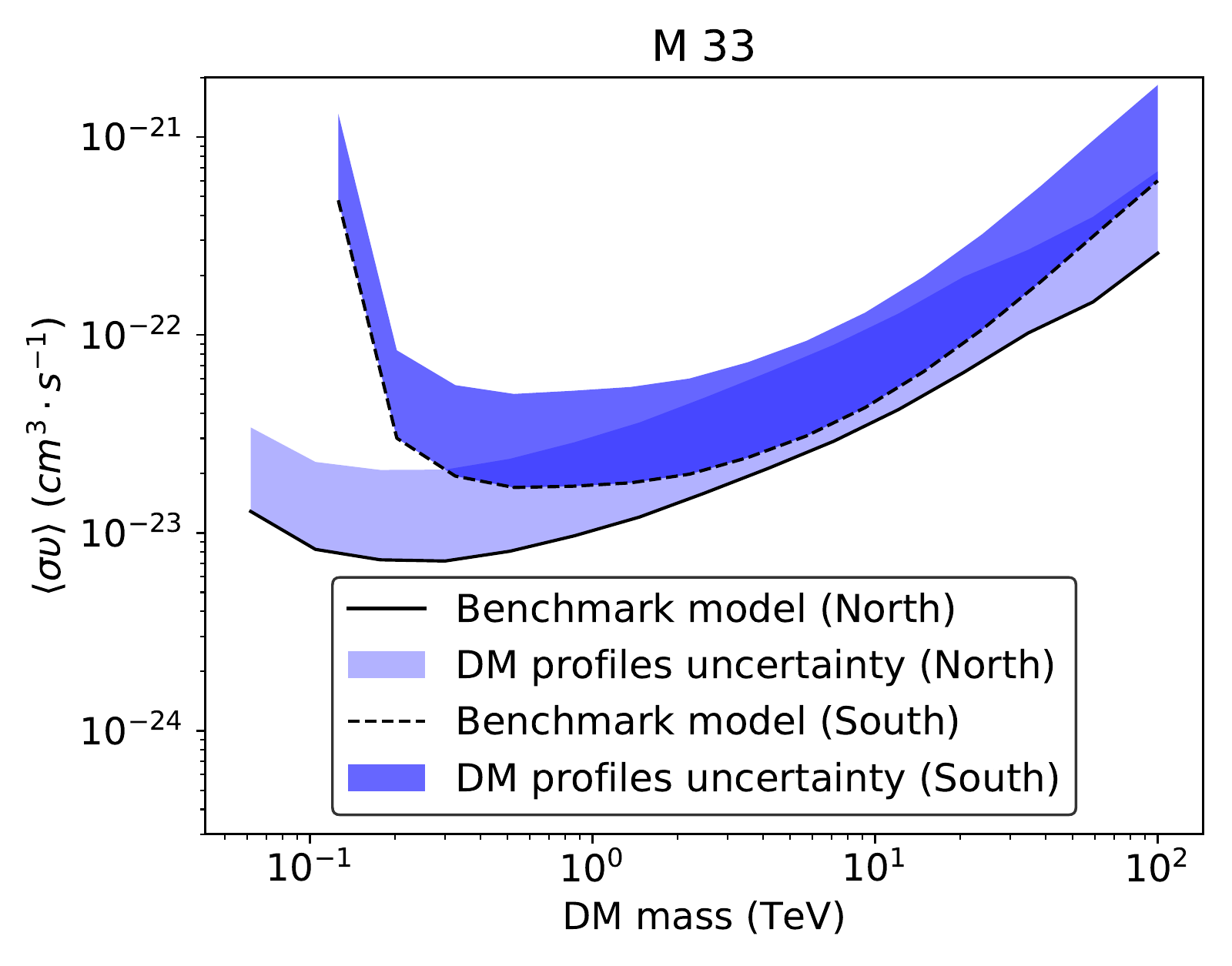}
    \caption{ Left: M31 upper limits uncertainty region for all different DM profiles -- $\tau^{+}\tau^{-}$ annihilation channel. With the black solid line, we highlight the upper limits for the benchmark model. Right: M33 upper limits uncertainty region for all different DM profiles -- $\tau^{+}\tau^{-}$ annihilation channel. With the black solid/dashed line, we highlight the upper limits for the benchmark model for the Northern/Southern CTA site respectively.}
    \label{TT}
    
\end{figure*}

\begin{figure*}[!h]
    \centering
    \includegraphics[width=0.497\textwidth]{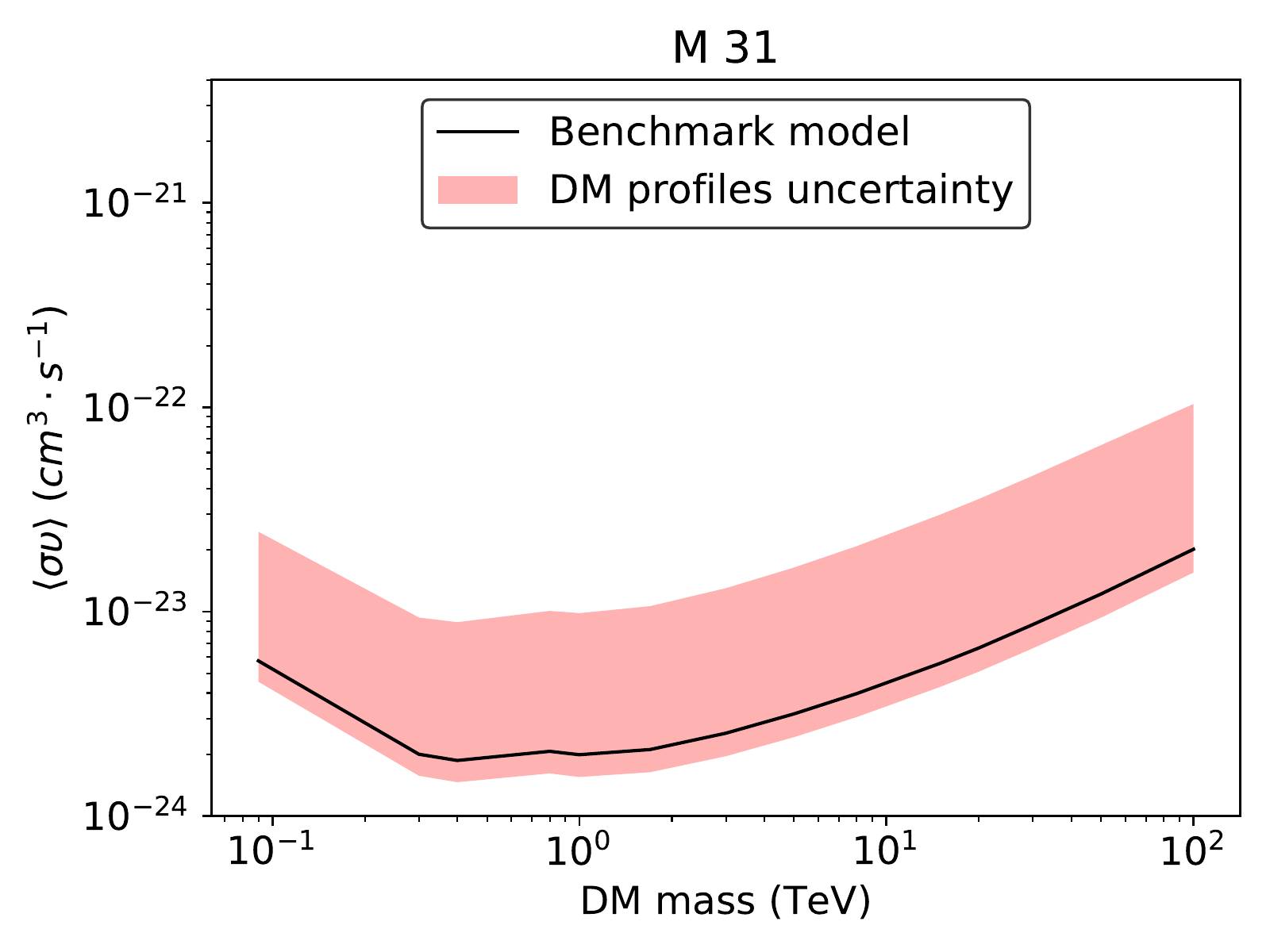}
    \includegraphics[width=0.495\textwidth]{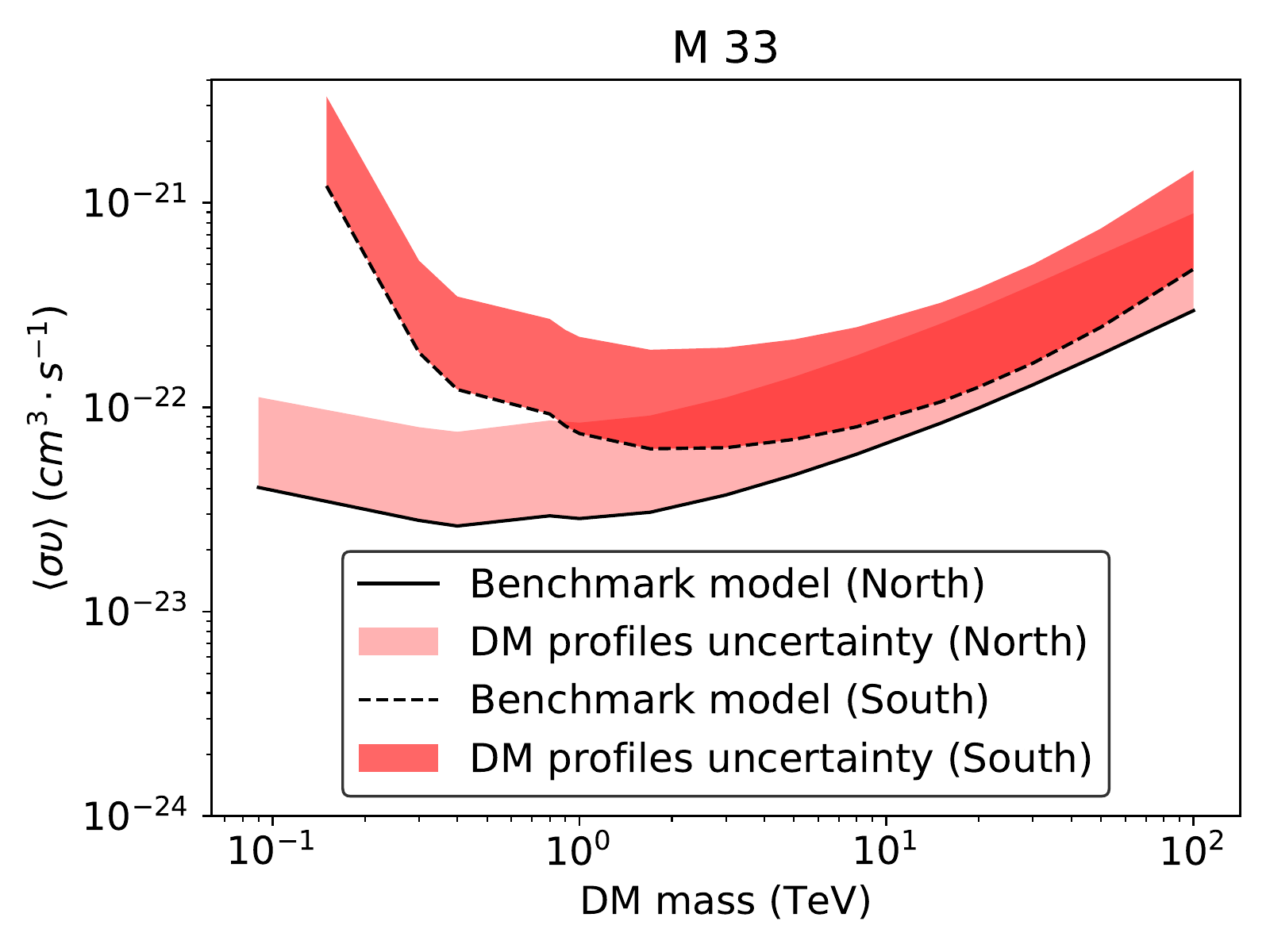}
    \caption{Left: M 31 upper limits uncertainty region for all different DM profiles -- $W^{+}W^{-}$ annihilation channel. With the black solid line, we highlight the upper limits for the benchmark model. Right: M 33 upper limits uncertainty region for all different DM profiles -- $W^{+}W^{-}$ annihilation channel. With the black solid/dashed line, we highlight the upper limits for the benchmark model for the Northern/Southern CTA site respectively.}
    \label{WW}
    
\end{figure*}

Using the profiles reported in Tab~\ref{tab:profiles} and \ref{tab:profiles2} we derived, additionally to Fig~\ref{BB} --  representing upper limits towards the benchmark annihilation channel -- the $95~\%$  confidence level upper limits for DM annihilation in the direction of both M31 and M33 for the rest two ($\tau^{+}\tau^{-}$ and $W^{+}W^{-}$) representative channels for DM searches. The obtained results are presented in Fig~\ref{TT} and \ref{WW}.

\section{Astrophysical emitting gamma-ray sources}\label{appendD}
The astrophysical sources within $5^\circ$ from the positions of M31 and M33 detected in the GeV band are summarized in Tab~\ref{ALL-SOURCES}. The point sources are adapted from 3FHL catalogue~\citep{Ajello_2017} of \flat sources detected above 10~GeV, the parameters of diffuse source (``Inner M31'') are adapted from \citet{2019ICRC...36..570K}. The table summarizes basic information about the sources (catalogue/reference, coordinates, suggested in 3FHL type and redshift) as well as spectral parameters of sources in the GeV band (spectral shape, slope, and flux).

\begin{table}
 \begin{adjustwidth}{-1.9cm}{}
\begin{tabular}{ |c|c|c|c|c|c|c|c|c| } 
\hline
 \multicolumn{9}{|c|}{Point sources} \\
 \hline Catalog & Source name & RA$~\degree$ &Dec$~\degree$&Class&z& spectral shape & Index  & \thead{Integrated Flux\\ $\mathrm{10^{-11}~ph~cm^{-2}~s^{-1}}$}  \\
 \hline
 \multicolumn{9}{|c|}{M31} \\
 \hline 3FHL & J0055.8+4507 & 13.95 & 45.13 & -&- & PowerLaw & -3.47  & $2.11$\\ 
 \hline
 3FHL &  J0039.2+4330 &   9.81 &   43.51 &bcu   & - &    Powerlaw   & -4.11   & $2.38$\\
 \hline
3FHL    &	J0049.0+4224 &    12.27  & 42.40 & -& - &     Powerlaw   & -2.33    & $2.65$ \\
 \hline
3FHL    &	J0040.3+4049 &    10.09  & 40.83&bcu &  - &     Powerlaw   &  -1.56    &  $2.48$ \\
 \hline
3FHL    &	J0047.9+3947 &    11.98  & 39.79 &  bll&0.25 &     Powerlaw   &  -2.33   &   $8.37$\\
 \hline
3FHL    &	J0041.5+3759 &    10.38  & 37.99 &  bcu&0.38 &     Powerlaw   &  -1.86   &  $2.82$\\
 \hline
 \multicolumn{9}{|c|}{M33} \\
 \hline
 3FHL & J0123.0+3422 & 20.77 & 34.37 & bll&0.27 & PowerLaw & -2.03  & $6.24$\\ 
 \hline
 3FHL & J0112.9+3208 & 18.24  &   32.15    &fsrq& 0.60 &    Powerlaw   & -2.69  & $3.57$\\
 \hline
3FHL    &	J0134.4+2638 &    23.61  & 26.65 & bcu &- &     Powerlaw   & -2.17    & $7.79$ \\
 \hline
3FHL    &	J0144.5+2705 &    26.14  & 27.09 & bll & - &     Powerlaw   &  -2.88    &  $25.19$ \\
 \hline
 \multicolumn{9}{|c|}{Extended source} \\
 \hline Reference & Source name & RA$~\degree$&Dec$~\degree$& spatial shape&Size& spectral shape & Index  &  \thead{Integrated Flux\\ $\mathrm{10^{-9}~ph~cm^{-2}~s^{-1}}$} \\
 \hline
 \citep{2019ICRC...36..570K} & Inner M31 & 10.68 & 41.26& Radial disk&$0.4\degree$& Powerlaw& $-2.8\pm0.3$& $0.5$\\
 \hline
 
\end{tabular}
\end{adjustwidth}
\caption{GeV/TeV sources, within $5\degree$ radius (CTA FoV) from M31 and M33 detected by Fermi/LAT above 30~GeV. The first four columns stand for the reference/catalogue, sources' names, and coordinates. The fifth column stands for the class (as indicated in 3FHL catalogue) of the point sources or the spatial shape of the Inner M31, where bll corresponds to Bl Lac balzars, bcu -- balzars of uncertain type, and the frsq  -- flat spectrum radio quasars. The sixth column indicates the redshift of the point sources and the spatial size of the Inner M31. The next two columns stand for the spectrum (spectral shape and index) for all sources. Finally, the last columns report the total integrated flux in 10-1000~GeV range for the point sources and in 1-100~GeV range for the Inner M31.}

\label{ALL-SOURCES}
\end{table}

\section{Contribution of the galactic diffuse halo}\label{appendA}

The expected flux from DM self-annihilation is proportional to the square of the DM density integrated along the line of sight (see J-factor Eq~\ref{eq:j_factor}). Calculating the expected flux from DM annihilation, one should take into account the contribution from DM annihilation signal originating from the MW DM halo. The $J$-factor of the MW halo is given by:
\begin{equation*}
      J(\psi)    =  \int\limits^{\ell_{max}}_{0}\rho^{2}(\sqrt{R^{2}_{sc}-2\ell R_{sc}\cos{\psi}+\ell^{2}}) d\ell
    \label{eq10}
\end{equation*}
where $\psi$ is the angular distance from the GC, $R_{sc}=8.5~\mathrm{kpc}$ is the Sun -- Galactic Center distance and $l_{max}$ is defined as
\begin{equation*}
      \ell_{max}
    = \sqrt{R^{2}_{MW}+R^{2}_{sc}\sin^{2}{\psi}}+R_{sc}\cos{\psi} \vspace{0.5cm}
    \label{eq11}
\end{equation*}
where $R_{MW}$ corresponds to the radius of MW DM halo. In this analysis, we consider $R_{MW}=\infty$, which results in $\ell_{max}=\infty$, since the contribution of the signal at large radii is negligible in comparison to the signal closer to the center.

The list of DM density profiles in the MW existing in the literature is given in Tab~\ref{tab:profiles1}~\citep[see] [and references therein]{2011PhRvD..83b3518P,2018PhRvL.120t1101A}.
In what below, we briefly summarise the profiles present in the table and which are characterised by four parameters ($\alpha,~\beta,~ \gamma,~ \delta$): a generalized profile proposed by \citet{1990ApJ...356..359H}, \citet{1993MNRAS.265..250D} and \citet{1996MNRAS.278..488Z}. Different combinations of the four parameter values lead to different DM distribution, i.e., (1, 3, 1, 0) corresponds to the widely used NFW profile \citep{1997ApJ...490..493N}, (2, 3, 1, 1) corresponds to a Burkert profile \citep{1995ApJ...447L..25B}, (1.5, 3, 1.5, 0) corresponds to a Moore profile and (2, 2, 0, 0) to an Isothermal profile \citep{1966AJ.....71...64K}. Einasto profile \citep{1965TrAlm...5...87E} follows a different parametrization based on a single parameter $\alpha$. A different parametrization of a DM density profile based on five different parameters ($r_{o},~a,~ \alpha,~\beta,~ \gamma$) is also found in \citet{2007PhRvD..76f3006P}; following the equations:
\
\begin{equation*}
\large{
    \begin{array}{rrcl}
        Hernquist: &\rho_{Her}(r)& =&\rho_s\cdot(\delta+\frac{r}{r_s})^{-\gamma}\cdot(1+(\frac{r}{r_s})^{\alpha})^{\frac{\gamma-\beta}{a}}\vspace{0.5cm}\\
        
        Einasto:&\rho_{Ein}(r)&=&\rho_s\cdot exp\{-\frac{2}{\alpha}\cdot[(\frac{r}{r_s})^{\alpha}-1]\}\vspace{0.5cm}\\
        
        Pullen:&\rho_{Pul}(r)& =&\rho_s\cdot(\frac{r_{o}}{r})^{\gamma}\cdot\frac{[1+(\frac{r_{o}}{r_s})^{\alpha}]^{\frac{\beta-\gamma}{\alpha}}}{[1+(\frac{r}{r_s})^{\alpha}]^{\frac{\beta-\gamma}{\alpha}}}\\
        
    \end{array}
}
\label{eq5}
\end{equation*}
where $r_s$ is the scale radius and $\rho_s$ is the scale density of the profile. For the DM profile presented in \citet{2007PhRvD..76f3006P},  $r_{o}=8.5$~kpc is the distance from the Sun to the GC.

Some other profiles can be obtained as the combination of two or more of the profiles above. A characteristic example is the HYB profile, which is a combination of SIS and NFW (hereafter hybrid profile).

In Tab~\ref{tab:profiles1} all Einasto profiles use $\alpha=0.17$ except for Einasto* which uses $\alpha=0.22$.
The majority of the remaining profiles are described by the generalized profile proposed by \citet{1990ApJ...356..359H}, \citet{1993MNRAS.265..250D} and \citet{1996MNRAS.278..488Z} or the Pullen parametrization profiles (see Eq~\ref{eq5}).

\begin{table}[]
    \centering
    \begin{adjustwidth}{-1.2cm}{}
    \begin{tabular}{|c|c|c|c|c|c|c|c|}
    \hline \hline
        \textbf{Profile} & \textbf{$\alpha$} & \textbf{$\beta$} & \textbf{$\gamma$} & \textbf{$\delta$} & \textbf{$r_s$}  & \textbf{$\rho_s$}  & \textbf{Reference} \\
         & & & & &\textbf{kpc}  & \textbf{GeV/cm$^3$}  \\
        \hline
        NFW(VLII) & $1$ & $3$ & $1$ & $0$ & $21$ & $0.307\cdot10^{0}$  & \citet{2011PhRvD..83b3518P} \\
        NFW &  $1$ & $3$ & $1$ & $0$ & $21$ & $0.307\cdot10^{0}$ &  \citet{2018PhRvL.120t1101A}\\
        NFW  & $1$ & $3$ & $1$ & $0$ &$16.1^{17}_{-7.8}$ & $0.531\cdot10^{0}$ &  \citet{2013JCAP...07..016N,2015EPJC...75..492A} \\
        NFW &  $1$ & $3$ & $1$ & $0$ & $20$ & $0.259\cdot10^{0}$ &  \citet{2011PhRvD..84b2004A}\\
        NFW &  $1$ & $3$ & $1$ & $-$ & $25$ & $0.3\cdot10^{0}$ &  \citet{2007PhRvD..76f3006P}\\
        NFW & $1$ & $3$ & $1$ & $0$ & $21.7$ & $0.303\cdot10^{0}$ &  \citet{2015JCAP...10..068A}\\
        NFW-c & - & - & $1.2$ & - & $21.7$ & $0.207\cdot10^{0}$ &  \citet{2015JCAP...10..068A}\\
        NFW-c &  - & - & $1.3$ & - & $20$ & $0.271\cdot10^{0}$ &  \citet{2015JCAP...05..011A}\\
        NFW &  $1$ & $3$ & $1$ & $0$ & $20$ & $0.345\cdot10^{0}$ &  \citet{2017PhRvD..95j3005K}\\
        NFW-c & - & - & $1.2$ & - & $20$ & $0.271\cdot10^{0}$ &  \citet{2017PhRvD..95j3005K}\\
        NFW & $1$ & $3$ & $1$ & $0$ & $20$ & $0.345\cdot10^{0}$ &  \citet{2016PDU....12....1D}\\
        NFW-c & - & - & $1.2$ & - & $20$ & $0.271\cdot10^{0}$ &  \citet{2016PDU....12....1D}\\
        NFW-c & - & - & $1.4$ & $0$ & $20$ & $0.213\cdot10^{0}$ &  \citet{2016PDU....12....1D}\\
        NFW &  $1$ & $3$ & $1$ & $0$ & $23.8$ & $0.14\cdot10^{0}$ &  \citet{2013JCAP...10..029G}\\
        NFW-c & $0.76$ & $3.3$ & $1.37$ & $0$ & $18.5$ & $0.23\cdot10^{0}$ &  \citet{2013JCAP...10..029G}\\
        \hline
        Einasto(Aq) & $0.17$ & - & -& - & $20$ & $0.106\cdot10^{0}$ & \citet{2011PhRvD..83b3518P} \\
        Einasto & $0.17$ & - & -& - & $20$ & $0.079\cdot10^{0}$ &  \citet{2018PhRvL.120t1101A} \\
        Einasto & $0.17$ & - & -& - & $28.4$ & $0.033\cdot10^{0}$ &  \citet{2018PhRvL.120t1101A} \\
        Einasto & $0.16$ & - & -& - & $20$ & $0.0606\cdot10^{0}$ &    \citet{2011PhRvD..84b2004A} \\
        Einasto & $0.17$ & - & -& - & $21.7$ & $0.0707\cdot10^{0}$ &    \citet{2015JCAP...10..068A} \\
        Einasto & $0.17$ & - & -& - &$20$ & $0.081\cdot10^{0}$ &    \citet{2016PDU....12....1D} \\
        Einasto* & $0.22$ & - & -&- & $19.7$ & $0.08\cdot10^{0}$ &    \citet{2013JCAP...10..029G} \\
        \hline
        Burkert & $2$ & $3$ & $1$ & $1$ & $9.26^{5.6}_{-4.2}$ & $1.568\cdot10^{0}$ & \citet{2013JCAP...07..016N,2015EPJC...75..492A}\\
        Burkert & $2$ & $3$ & $1$ & $1$ & $2$ & $37.76\cdot10^{0}$ & \citet{2013JCAP...10..029G}\\
        \hline
        Moore & $1.5$ & $3$ & $1.5$ & $0$ & $28$ & $0.0527\cdot10^{0}$ & \citet{2011PhRvD..84b2004A}\\
        \hline
        Kravtsov & $2$ & $3$ & $0.4$ & $0$ & $10$ & $0.703\cdot10^{0}$ & \citet{2011PhRvD..84b2004A}\\
        \hline
        Isothermal & $2$ & $2$ & $0$ & $-$ & $4$ & $0.3\cdot10^{0}$ & \citet{2007PhRvD..76f3006P}\\
        Isothermal & $2$ & $2$ & $0$ & $0$ & $4$ & $2.206\cdot10^{0}$ & \citet{2015JCAP...10..068A}\\
        \hline
        Ka & $2$ & $3$ & $0.2$ & $-$ & $11$ & $0.4\cdot10^{0}$ & \citet{2007PhRvD..76f3006P}\\
        Kb & $2$ & $3$ & $0.4$ & $-$ & $12$ & $0.4\cdot10^{0}$ & \citet{2007PhRvD..76f3006P}\\
        \hline\hline
    \end{tabular}
    \end{adjustwidth}
    \caption{distribution profiles in GC}
    \label{tab:profiles1}
\end{table}

In the case of the Burkert profile in \citet{2013JCAP...10..029G}, we choose $r_{s}=2~\mathrm{kpc}$ based on \citep{Hooper_2013,2011ApJ...742...76G,2012ApJ...761...91A} and using, for the normalization, the local density suggested in \citet{2010JCAP...08..004C}. This value appears to be compatible with the observational constraints from \citet{2011JCAP...11..029I}. However, a more recent work favors a much larger scale radius and a slightly different normalization for Burkert profiles (see \citet{2013JCAP...07..016N} and \citet{2015EPJC...75..492A}).

\begin{figure*}[!h]
    \centering
    \includegraphics[width=0.514\textwidth]{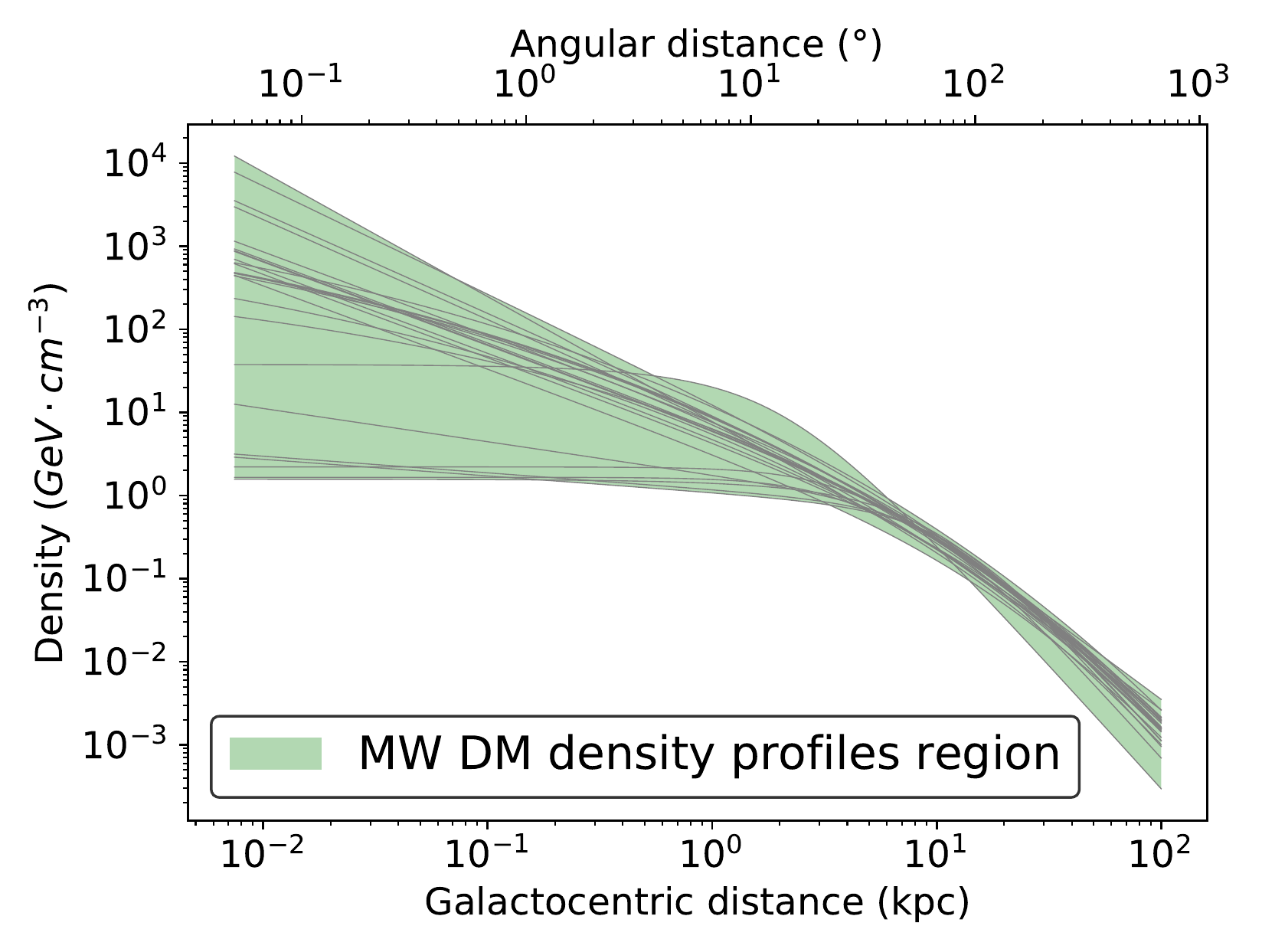}
    \includegraphics[width=0.478\textwidth]{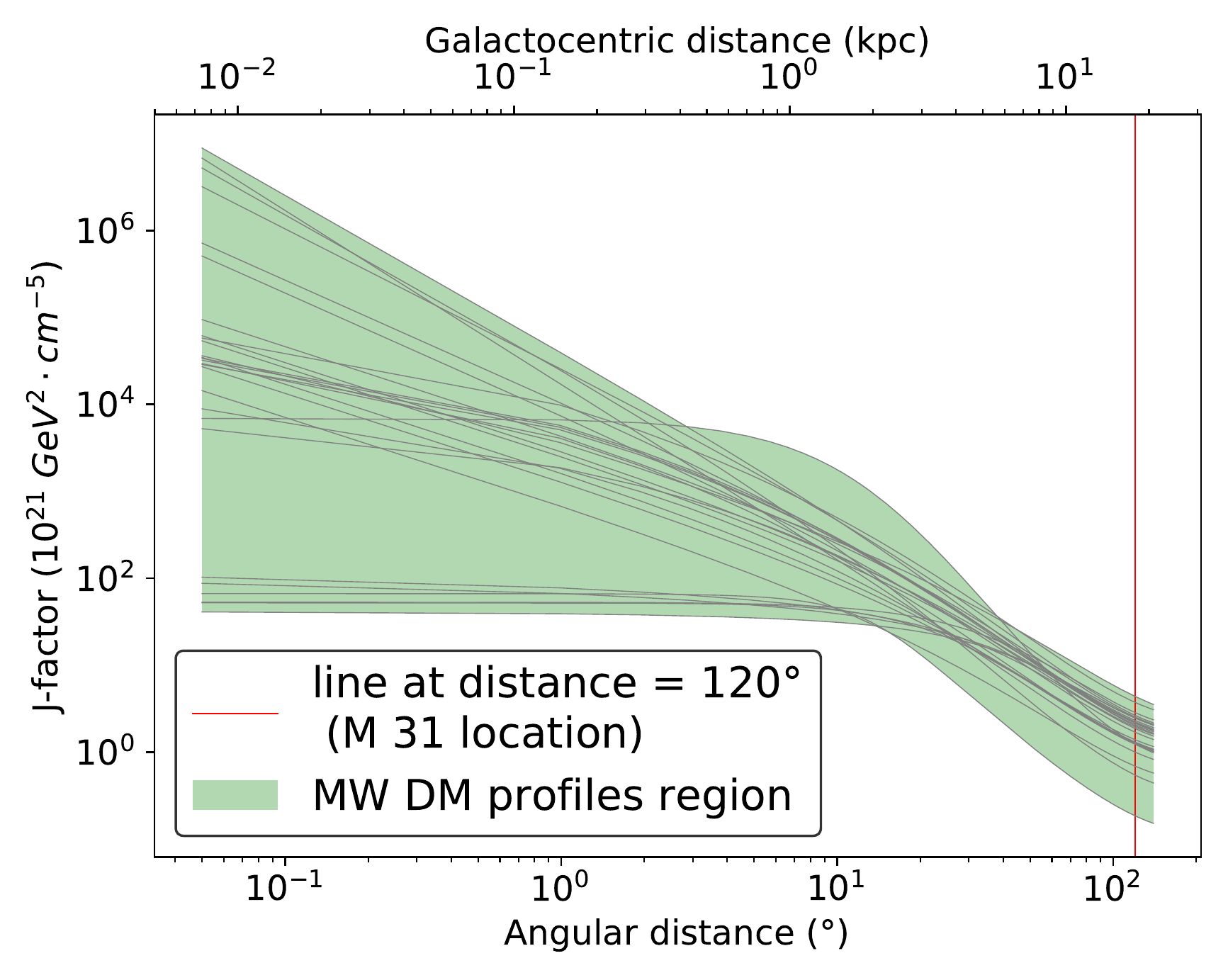}
    
    \caption[DM density profiles and J-factor values of the GC] {Left panel: DM density profiles (see Tab~\ref{tab:profiles1}) in our galaxy as a function of distance from the GC, in kpc. Right panel: $J$-factor plotted as a function of angular distance in degrees from the GC, for all the different profiles (see Tab~\ref{tab:profiles1}). In the plot there is a vertical line, in red, which corresponds to the angular distance of M31 from the GC.}
    \label{GC}
    
\end{figure*}

\begin{figure*}[!h]
    \centering
    \includegraphics[width=0.8\textwidth]{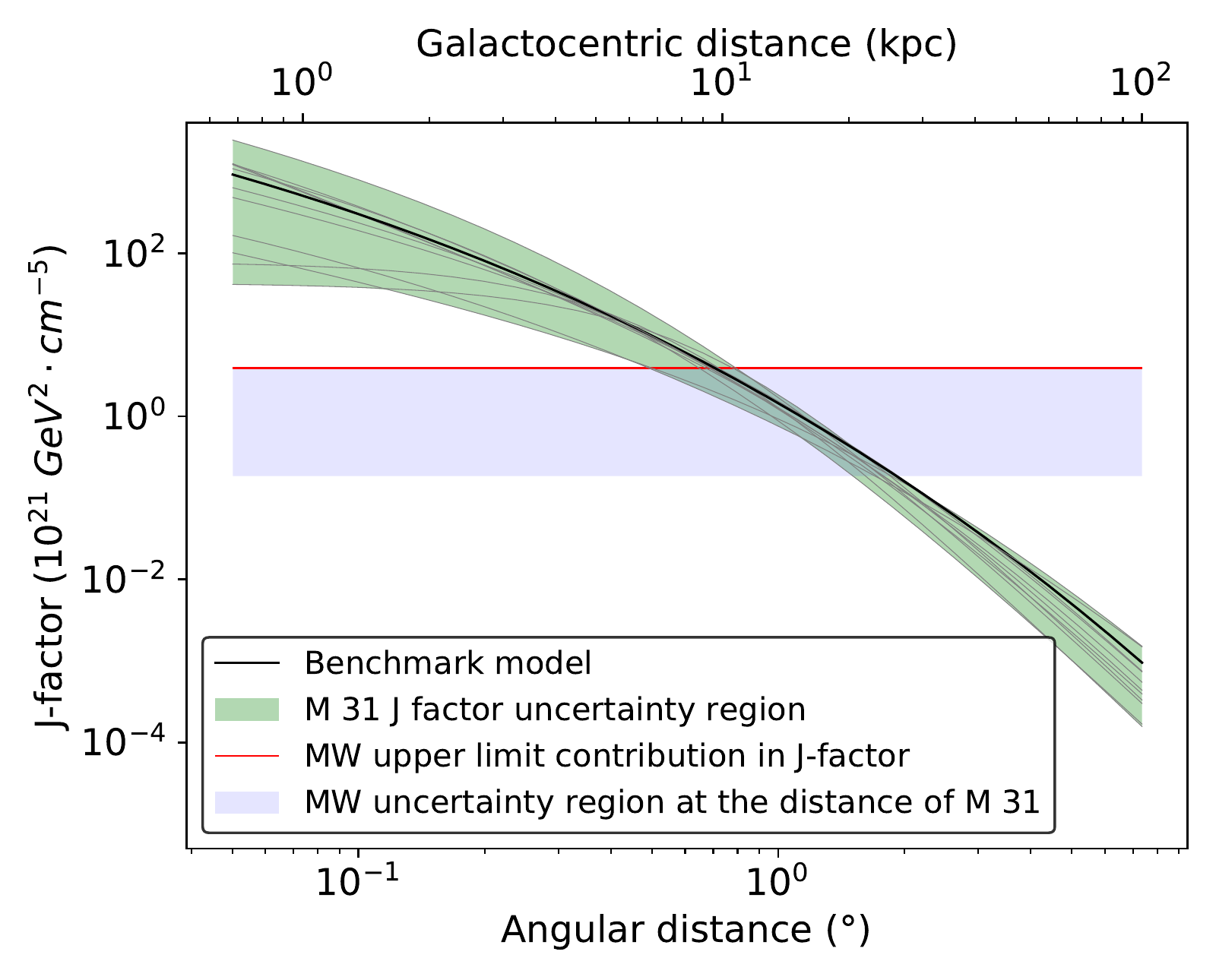}
    \caption[$J$-factors comparison for M31 and MW at the position of M31] {$J$-factors for M31 and the GC. The green band is the J-factor uncertainty for M31 from this work. The solid black lines correspond to the twelve different DM profiles that we collected from the literature. The benchmark model is highlighted with the bold black solid line. The blue band is the J-factor uncertainty region as seen from the GC at the distance of M31 galaxy. The red solid line stands for the upper limits contribution of the MW to the J-factor values of M31.}
    \label{overplot}
    
\end{figure*}

In Fig~\ref{GC} we present the obtained profiles of the DM density distribution reported in Tab~\ref{tab:profiles1} and the corresponding J-factors.
Fig~\ref{overplot} shows the relative contribution of the MW DM halo in comparison to the signal from M31. The signal from the M31 center can be at least two orders of magnitude exceeding the MW DM halo contribution. The M31 and MW DM halo signals become equal only at about $1\degree$ away from M31 center. Being sub-dominant, the contribution from MW DM halo was neglected in this paper.

\section{Effect of the DM substructures to the upper limits results}\label{appendB}

As an additional step, we cross-checked the used DM signal templates with ones produced by \texttt{CLUMPY} software\footnote{\url{https://clumpy.gitlab.io/CLUMPY/}}. The utilisation of this software allowed us also to estimate the contribution from the DM substructures present in M31/M33 DM halo. 

Given that the total DM density distribution is the sum of a smooth contribution and a distribution of sub-halos, the latter must be interpreted as scaled-down versions of the host halo. The presence of such substructures can significantly enhance the expected signal, and therefore their implementation should be properly treated.

For the modeling of the substructures, we selected a substructure spatial distribution $dN_{sub}/dV$ that follows the smooth parent halo profile. A mass density distribution described by the function $dN_{sub}/dM\propto M^{-\alpha_{M}}$, with $\alpha_{M}=1.9$ and $10\%$ mass fraction in substructures, was considered as suggested by numerical simulations of Milky-like halos \citep{2008MNRAS.391.1685S,2008ApJ...679.1260M}. The threshold mass for the smallest and the most massive subhalos are fixed to $10^{-6}~\mathrm{M\odot}$ and $10^{-2}~\mathrm{M_{tot}}$ respectively, when $\mathrm{M_{tot}}$ is the total mass of the corresponding galaxy, utilizing the subclumps mass-concentration relationship reported in \citet{2014MNRAS.442.2271S}. However, there are several works that suggest that the concentration of subhalos is greater in comparison to that of field halos of the same mass, which indicates a larger substructure boosting factor \citep{2000ApJ...544..616G,2001MNRAS.321..559B,2007ApJ...667..859D,2008Natur.454..735D,2015PhRvD..92l3508B,2016MNRAS.457..986Z}. It is noteworthy, that \citet{2017MNRAS.466.4974M} attempted to refine the substructure boost model provided by \citet{2014MNRAS.442.2271S} by utilizing data of N-body \texttt{Via Lactea} and \texttt{Elvis} Milky Way size-simulations. They obtained boost values of a factor of 2-3 greater in comparison to previous reports. However, one has also to consider the suppression level on the boosting factor when considering unavoidable tidal stripping effects - which appears to suppress significantly the boost factor in cases of dSphs subhalos (only a few tens of percent gain on the total boost factor is obtained in such cases) whereas it introduces an intermediate suppressing of the total boost factor for field halos such as M31 (of the level of $20-30\%$). In this case, we would expect moderately more constraining upper limits results in comparison to those obtained in this work by adopting the more conservative benchmark boost factor introduced in \citet{2014MNRAS.442.2271S}. A more complete overview of the impact of the substructure boosting on the upper limit results on the DM annihilation cross-section from extragalactic halo observations is given in \citet{2019Galax...7...68A} where the authors acknowledge that numerical simulations provide the most accurate assessment in resolved regimes, however, they pinpoint the dangers of the unavoidable extrapolation of the substructure properties which introduces large uncertainties to the heavily enhanced obtained boost factor, and thus such results should be treated with caution. Such a high uncertainty on the obtained boosting factor becomes evident when nearly every single individual work reports on a different derived boosting factor, ranging from 2 to values greater than 100 for galaxy-size halos. As a complementary approach, they provide great insight into semi-analytic modelings, such as Press-Schechter formalism and tidal-stripping modeling, which in contrast to N-body simulations appear to be more modest resulting in an order of unity for galaxy-size halos. One could even consider a much greater enhancement on the substructure boosting when considering that prompt DM cusps survived tidal stripping and thus are present today, as introduced in \citet{2022arXiv220911237D}. To summarize, computing the exact boost factor that DM subhalos introduce; comes as a great challenge, and it is still remaining highly uncertain; thus in this work, we adopt the most conservative approach introduced in \citet{2014MNRAS.442.2271S} aiming at not overestimating our upper limits result.

\begin{figure*}[h!]
    \centering
    \includegraphics[width=0.8\textwidth]{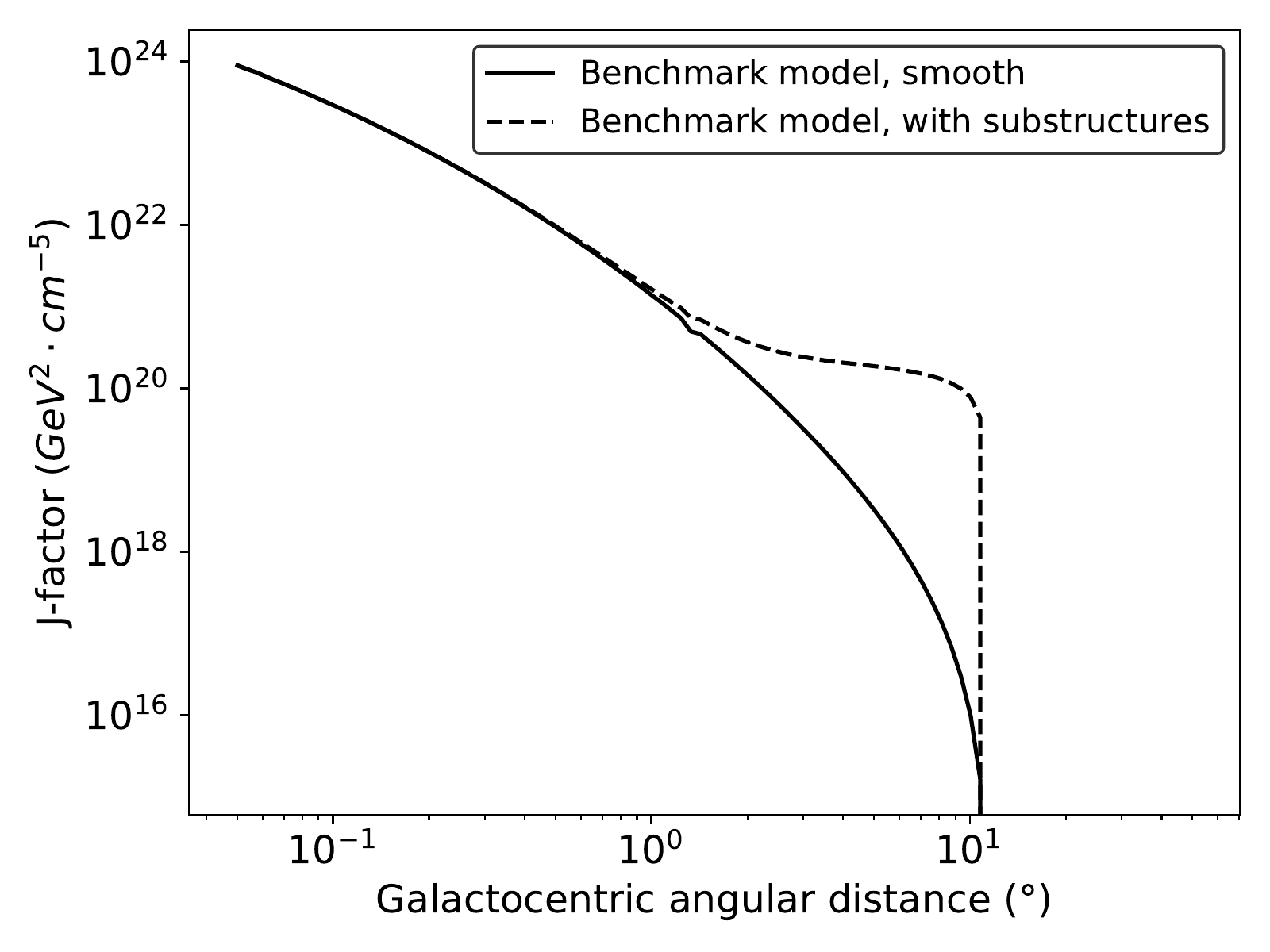}
    \caption[J-factor comparison with and without substructures]{J-factor as a function of the angular distance for the benchmark profile of M31 galaxy. The black solid line corresponds to the J-factor values without including substructures. The black dashed line corresponds to the J-factor values when including substructures.}
    \label{J-SUB}
    
\end{figure*}

Fig~\ref{J-SUB} shows the radial dependency of the J-factor, as obtained using \texttt{CLUMPY} v3.0.1 code \citep{2012CoPhC.183..656C,BONNIVARD2016336,2019CoPhC.235..336H}, for the benchmark profile of M31 that was considered in this work, verifying the significant contribution of the sub-halos at the outskirts of the parent halo. The above behavior has been analytically discussed in \citet{Han_2016}, and it is attributed to the decreasement of the fraction of the mass bound of the substructures towards the center of the galaxy due to tidal stripping.

\begin{figure}[h!]
    \centering
    \includegraphics[width=0.497\textwidth]{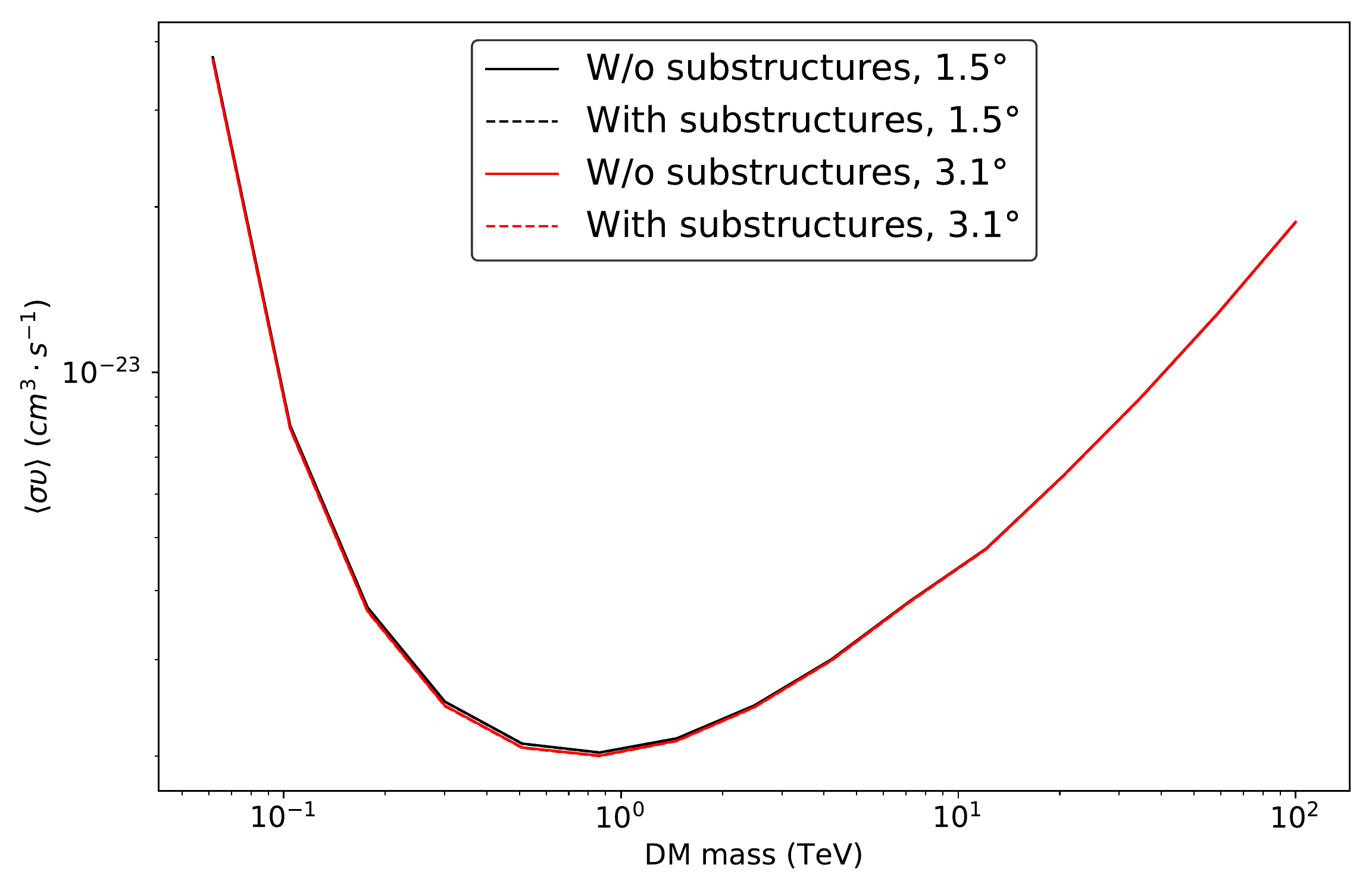}
    \includegraphics[width=0.495\textwidth]{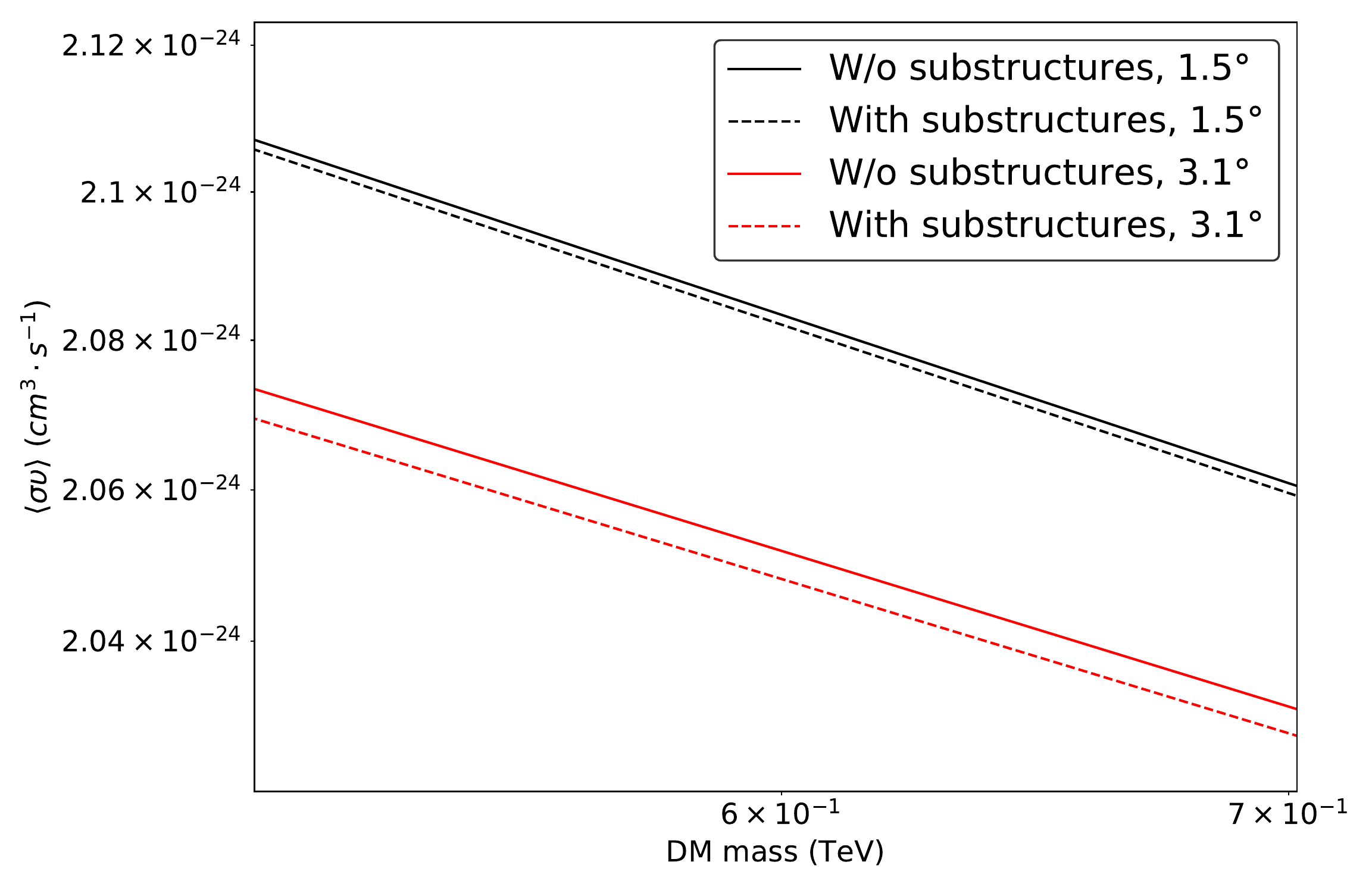}
    \caption[$\langle\sigma\upsilon\rangle$ limits ($2\sigma$) when including substructures for the benchmark model of M31; using prod3b-v2 IRF]{M31 upper limits ($2\sigma$) comparison for the benchmark model, with and without substructures. Left plot: The solid lines (both black and red) correspond to the benchmark DM profile without the presence of substructures, when considering $1.5\degree$ and $3.1\degree$ spatial size of the DM source template respectively. The dashed lines (both the black and the red one) correspond to the benchmark DM profile in the presence of substructures, when considering $1.5\degree$ and $3.1\degree$ spatial size of the DM source template respectively. Right plot: Left plot zoomed in for higher accuracy.}
    \label{SUBS}
    
\end{figure}

In Fig~\ref{SUBS} we present the more constraining results that we obtain in the presence of DM substructures in comparison to the smooth profile that we considered as benchmark profile for M31 in this work. In addition, this figure shows that when considering a larger DM source template, where the subhalos contribution is stronger, the upper limits results become even more constraining in comparison to a smaller DM source template.

The upper limit results when including substructures should not be considered final certain results in any case, since the nature of the DM sub-halos is still unrevealed. When more information and details for the actual nature of those substructures (i.e. mass, spatial distribution as well as the description of the DM distribution within each halo; changing each of those parameters results in different J-factor values) will be available, then more accurate analysis will be conducted.

\end{appendices}

\def\aj{AJ}%
\def\actaa{Acta Astron.}%
\def\araa{ARA\&A}%
\def\apj{ApJ}%
\def\apjl{ApJ}%
\def\apjs{ApJS}%
\def\ao{Appl.~Opt.}%
\def\apss{Ap\&SS}%
\def\aap{A\&A}%
\def\aapr{A\&A~Rev.}%
\def\aaps{A\&AS}%
\def\azh{AZh}%
\def\baas{BAAS}%
\def\bac{Bull. astr. Inst. Czechosl.}%
\def\caa{Chinese Astron. Astrophys.}%
\def\cjaa{Chinese J. Astron. Astrophys.}%
\def\icarus{Icarus}%
\def\jcap{J. Cosmology Astropart. Phys.}%
\def\jrasc{JRASC}%
\def\mnras{MNRAS}%
\def\memras{MmRAS}%
\def\na{New A}%
\def\nar{New A Rev.}%
\def\pasa{PASA}%
\def\pra{Phys.~Rev.~A}%
\def\prb{Phys.~Rev.~B}%
\def\prc{Phys.~Rev.~C}%
\def\prd{Phys.~Rev.~D}%
\def\pre{Phys.~Rev.~E}%
\def\prl{Phys.~Rev.~Lett.}%
\def\pasp{PASP}%
\def\pasj{PASJ}%
\def\qjras{QJRAS}%
\def\rmxaa{Rev. Mexicana Astron. Astrofis.}%
\def\skytel{S\&T}%
\def\solphys{Sol.~Phys.}%
\def\sovast{Soviet~Ast.}%
\def\ssr{Space~Sci.~Rev.}%
\def\zap{ZAp}%
\def\nat{Nature}%
\def\iaucirc{IAU~Circ.}%
\def\aplett{Astrophys.~Lett.}%
\def\apspr{Astrophys.~Space~Phys.~Res.}%
\def\bain{Bull.~Astron.~Inst.~Netherlands}%
\def\fcp{Fund.~Cosmic~Phys.}%
\def\gca{Geochim.~Cosmochim.~Acta}%
\def\grl{Geophys.~Res.~Lett.}%
\def\jcp{J.~Chem.~Phys.}%
\def\jgr{J.~Geophys.~Res.}%
\def\jqsrt{J.~Quant.~Spec.~Radiat.~Transf.}%
\def\memsai{Mem.~Soc.~Astron.~Italiana}%
\def\nphysa{Nucl.~Phys.~A}%
\def\physrep{Phys.~Rep.}%
\def\physscr{Phys.~Scr}%
\def\planss{Planet.~Space~Sci.}%
\def\procspie{Proc.~SPIE}%
\let\astap=\aap
\let\apjlett=\apjl
\let\apjsupp=\apjs
\let\applopt=\ao

\bibliographystyle{mnras}
\bibliography{biblio2.bib}

\end{document}